\documentclass[a4paper,12pt]{article}
\pdfoutput=1
\usepackage{amssymb}
\usepackage{amsmath}

\usepackage{epsfig}
\usepackage{graphicx}
\usepackage{slashed}

\makeatletter

\@addtoreset{equation}{section}
\makeatother
\usepackage{amsmath}
\numberwithin{equation}{section}
\usepackage{amsfonts}
\usepackage{slashed}

\def\be{\begin{equation}}
\def\ee{\end{equation}}
\def\bea{\begin{eqnarray}}
\def\eea{\end{eqnarray}}

\def\({\left(}
\def\){\right)}
\def\<{\left<}
\def\>{\right>}

\def\be{\begin{equation}}
\def\ee{\end{equation}}

\def\bea{\begin{eqnarray*}}
\def\eea{\end{eqnarray*}}

\def\ben{\begin{eqnarray}}
\def\een{\end{eqnarray}}

\def\({\left(}
\def\){\right)}
\def\<{\left<}
\def\>{\right>}
\def\!{\right|}
\def\|{\left|}

\def\[{\left[}
\def\]{\right]}

\def\+{\bar}
\def\mb{\mathbb}

\def\D{{\cal{D}}}
\def\L{{\cal{L}}}
\def\K{{\cal{K}}}
\def\P{{\cal{P}}}

\def\t{\widetilde}
\def\A{{\cal{A}}}

\def\g{{\mathfrak{g}}}

\def\N{{\cal{N}}}

\def\L{{\cal{L}}}

\def\l{\ell}

\def\h{\widehat}

\begin{document}

\setlength{\unitlength}{1mm}

\pagestyle{empty}
\vskip-10pt
\vskip-10pt
\hfill 
\begin{center}
\vskip 3truecm
{\Large \bf
The non-Abelian tensor multiplet}
\vskip 2truecm
{\large \bf
Andreas Gustavsson}
\vspace{1cm} 
\begin{center} 
Department of Physics and Astronomy, Uppsala University,\\
Box 516, SE-75120 Uppsala, Sweden
\end{center}
\vskip 0.7truecm
\begin{center}
(\tt agbrev@gmail.com)
\end{center}
\end{center}
\vskip 2truecm
{\abstract{We assume the existence of a background vector field that enables us to make an ansatz for the superconformal transformations for the non-Abelian 6d $(1,0)$ tensor multiplet. Closure of supersymmetry on generators of the conformal algebra, requires that the vector field is Abelian, has scaling dimension minus one and that the supersymmetry parameter as well as all the fields in the tensor multiplet have vanishing Lie derivatives along this vector field. We couple the tensor multiplet to a hypermultiplet and obtain superconformal transformations that we close off-shell.}}

\vfill
\vskip4pt
\eject
\pagestyle{plain}

\section{Introduction}
There might have been speculations that the non-Abelian 6d $(2,0)$ tensor multiplet theory could be a non-Abelian gerbe theory \cite{Witten:2009at}, although so far no successful such theory has been found. The simplest example of a non-Abelian gerbe is where one introduces a non-Abelian two-form together with a flat one-form gauge field \cite{Alvarez:1997ma}. Let us assume that a classical field theory description for the non-Abelian tensor multiplet exists, and that we just have not found it yet. Let us refer to it as the theory $X[G]$ where $G$ denotes the gauge group. The theory $X[G]$ might be characterized by a classical Lagrangian, or classical equations of motion, perhaps making use of some kind of non-Abelian gerbe construction. 

Not much may be known about the theory $X[G]$, but upon circle compactification and dimensional reduction down to 5d, it shall reduce to 5d MSYM with gauge group $G$.

If we just compactify one circle direction but do not perform the dimensional reduction, then there will appear modes that are the Kaluza-Klein (KK) modes if we rewrite the theory $X[G]$ in terms of 5d fields. The zeroth KK mode gives the 5d MSYM. 

The conjecture of \cite{Douglas:2010iu}, \cite{Lambert:2010iw} says that 6d $(2,0)$ theory compactified, but not dimensionally reduced, on a circle is actually equivalent with the 5d maximally supersymmetric Yang-Mills (MSYM) that one gets by performing dimensional reduction along that circle. 

But as we just compactify the classical field theory $X[G]$, we get all the higher KK modes in addition to the zero modes. We conclude that the two statements 
\begin{enumerate}
\item The conjecture of \cite{Douglas:2010iu}, \cite{Lambert:2010iw} holds.
\item The classical field theory $X[G]$ exists. 
\end{enumerate}
can not be simultaneously true. 

In this paper, we will assume that at least one of the two  statements above is correct. But as they can not both be correct, we see that either one of the two statements must be correct. But as we will argue, which one of the two statements 1 and 2 is realized, is not universal. It depends on the choice of gauge group $G$ among other things.

If the gauge group $G = U(1)$ is Abelian, then the theory $X[U(1)]$ exists. We have a 6d classical field theory description in terms of the 6d $(2,0)$ Abelian tensor multiplet fields. This means that the conjecture can not hold when $G=U(1)$, as was also pointed out in \cite{Dolan:2012wq}. 

We may deform the Abelian theory by making space noncommutative. This gives the Abelian 5d MSYM theory the structure reminiscent of a non-Abelian theory, due to a noncommutative star-product between the fields that live on a noncommutative space. In this deformed case, there does not seem to exist a classical field theory in 6d.\footnote{A no-go theorem to constructing a noncommutative M5 brane was found in \cite{Chen:2010ny}. This no-go theorem was circumvented in \cite{Gustavsson:2010nc}, but the star 3-product used there to circumvent the no-go theorem has no application to the noncommutative M5 brane. As the noncommutativity becomes small, the Abelian tensor multiplet theory becomes a good approximation. But it is not an exact description as long as the noncommutativity parameter is non-zero.} Indeed then it seems that the conjecture becomes true. It has been shown that the noncommutative 5d MSYM theory has noncommutative instanton particle solutions that capture all the missing KK modes \cite{Kim:2011mv}, \cite{Bak:2012ct} by comparing with Abelian 6d $(2,0)$ theory that should be a very good approximation for a very small noncommutativity parameter. 

We will assume that space is commutative and that the conjecture holds when the gauge group is non-Abelian and its Lie algebra is semi-simple so that it does not contain any free Abelian tensor multiplets. We will construct classical non-Abelian 6d field theories by following the constructions in \cite{Lambert:2010iw}, \cite{Chen:2013wya}. For these theories there is a vector field $v^M$ along which the fields have no spatial dependence. The Lie derivatives of all the classical fields vanish along this vector field. By allowing for such a vector field, we are able to construct 6d classical field theories with $(2,0)$ superconformal symmetry. In order for the vector field $v^M$ to not break conformal symmetry, it shall transform under conformal transformations like a spin-one field with scaling dimension $\Delta = -1$. The vector field must be an isometry, and if we compactify the 6d theory along those circles, we get 5d SYM theory with no additional KK modes because of the vanishing Lie derivatives along that vector field.

\section{Gauge symmetry}
We will not introduce a one-form gauge field $A_M$ as an independent field from the two-form $B_{MN}$. Instead we will follow \cite{Lambert:2010wm} and assume the existence of a vector field $v^M$. We will assume that $B_{MN}$ and $v^M$ both take values in a set $\A_3$ and that there exists a multiplication on $\A_3$ that we denote as $\<\bullet,\bullet\>$ that maps two elements of $\A_3$ into a Lie algebra $\g$ of some gauge group. This should be the gauge group that appears upon dimensional reduction on a circle down to 5d SYM. We define the $\g$-valued gauge field from $B_{MN}$ as 
\bea
A_M &=& \<B_{MN},v^N\>
\eea
For the gauge parameters, we assume the following relation,
\bea
\Lambda &=& \<\Lambda_M,v^M\>
\eea
between the $\g$-valued gauge parameter $\Lambda$, and the $\A_3$-valued gauge parameters $\Lambda_M$. If $\varphi, v^M\in \A_3$, then we assign the gauge variation rules
\bea
\delta \varphi &=& -(\Lambda,\varphi)\cr
\delta v^M &=& -(\Lambda,v^M)
\eea
where we use the round bracket to map elements in $\g\times \A_3$ to an element in $\A_3$. We also use the standard square bracket for the commutator of two elements in $\g$. If we multiply two fields $\varphi_1$ and $\varphi_2$ that are valued in $\A_3$, then we get a field that is valued in $\g$. Let us assume that this composite field is in the adjoint representation. Then we shall have the gauge variation
\bea
\delta\(\<\varphi_1,\varphi_2\>\) &=& - [\Lambda,\<\varphi_1,\varphi_2\>]
\eea
On the other hand, the gauge variation should satisfy the Leibniz rule,
\bea
\delta\(\<\varphi_1,\varphi_2\>\) &=& \<\delta \varphi_1,\varphi_2\> + \<\varphi_1,\delta \varphi_2\>
\eea
Putting these two ingredients together, we find the following Jacobi identity,
\bea
\<(\Lambda,\varphi_1),\varphi_2\> + \<\varphi_1,(\Lambda,\varphi_2)\> &=& [\Lambda,\<\varphi_1,\varphi_2\>]
\eea
We will take the product $\<\bullet,\bullet\>:\A_3 \times \A_3 \rightarrow \g$ to be antisymmetric. That antisymmetry does not follow from gauge symmetry, and indeed we will introduce another gauge invariant product that we will denote as $\bullet \cdot \bullet: \A_3 \times \A_3\rightarrow \mb{C}$ that is hermitian
\bea
\varphi_1 \cdot \varphi_2 &=& (\varphi_2 \cdot \varphi_1)^*
\eea
and gauge invariant
\bea
(\Lambda,\varphi_1)\cdot \varphi_2 + \varphi_1 \cdot (\Lambda,\varphi_2) &=& 0
\eea
This is the inner product that we will use for writing a gauge invariant action.

The infinitesimal gauge variation for $A_M$ is
\ben
\delta A_M &=& D_M \Lambda\label{gauge}
\een
The gauge covariant derivative is
\bea 
D_M \Lambda &=& \partial_M \Lambda + [A_M,\Lambda]\cr
D_M \varphi &=& \partial_M \varphi + (A_M,\varphi)
\eea
where it acts on a $\g$-valued field $\Lambda$ and on an $\A_3$ valued field $\varphi$ respectively. Let us expand (\ref{gauge}),
\bea
\delta A_M &=& \partial_M \<\Lambda_N,v^N\> + [A_M,\<\Lambda_N,v^N\>]\cr
&=& \<\partial_M \Lambda_N,v^N\> + \<\Lambda_N,\partial_M v^N\> + [\<B_{MN},v^N\>,\Lambda]\cr
&=& \<\partial_M \Lambda_N - \partial_N \Lambda_M,v^N\> +  [\<B_{MN},v^N\>,\Lambda]\cr
&& + \<\partial_N \Lambda_M,v^N\> + \<\Lambda_N,\partial_M v^N\>
\eea
We use
\bea
\delta A_M &=& \<\delta B_{MN},v^N\> + \<B_{MN},\delta v^N\>
\eea
and the Jacobi identity in the form
\bea
\<(\Lambda,v^N),B_{MN}\> - \<(\Lambda,B_{MN}),v^N\> &=& [\Lambda,\<v^N,B_{MN}\>]
\eea
Consistency requires that the first term $\<(\Lambda,v^N),B_{MN}\>$ is vanishing. But we do not want this to impose a constraint on $B_{MN}$ nor on $\Lambda_P$, but we want this to be a constraint on $v^M$ only. For this to be possible, we assume that we have associativity
\bea
(\<\Lambda_P,v^P\>,v^N) = \(\Lambda_P,\<v^P,v^N\>\)
\eea
As the product is associative, we can introduce a 3-bracket notation. For any three elements $a,b,c\in \A_3\times \A_3\times \A_3$, we define
\bea
[a,b,c] := (\<a,b\>,c) = (a,\<b,c\>)
\eea
We see that the consistency condition that we shall impose on $v^M$ is the Abelian constraint
\bea
\<v^M,v^N\> &=& 0
\eea
We also define
\bea
\<\L_v,\Lambda_M\> &:=&  \<v^N,\partial_N \Lambda_M\> + \<\partial_M v^N,\Lambda_N\>
\eea
Using this, we get
\bea
\delta A_M &=& \<\delta B_{MN},v^N\> + \<B_{MN},\delta v^N\> - \<\L_v,\Lambda_M\>
\eea
where
\bea
\delta B_{MN} &=& \partial_M \Lambda_N - \partial_N \Lambda_M - (\Lambda,B_{MN})\cr
\delta v^N &=& - (\Lambda,v^N)
\eea
We obtain the consistency condition
\bea
\<\L_v,\Lambda_M\> &=& 0
\eea

Let us move on to the field strength that we expand as
\bea
F_{MN} &=& \partial_M A_N - \partial_N A_M + [A_M,A_N]\cr
&=& \<\partial_M B_{NP},v^P\> + \<B_{NP},\partial_M v^P\>\cr
&& - \<\partial_N B_{MP},v^P\> - \<B_{MP},\partial_N v^P\>\cr
&& + [\<B_{MP},v^P\>,\<B_{NQ},v^Q\>]
\eea
Using the Jacobi identity
\bea
[A_M,\<B_{NP},v^P\>] &=& \<(A_M,B_{NP}),v^P\> - \<(A_M,v^P),B_{NP}\>
\eea
together with the fact that
\bea
(A_N,v^P) = [B_{NQ},v^Q,v^P] = 0
\eea
by the Abelian constraint on $v^M$, we get 
\bea
F_{MN} &=& \<\partial_M B_{NP} + \partial_N B_{PM} + \partial_P B_{MN},v^P\>\cr
&& - \<\partial_P B_{MN},v^P\> + \<B_{PN},\partial_M v^P\> + \<B_{MP},\partial_N v^P\>\cr
&& + \<(A_M,B_{NP}),v^P\>
\eea
If we note that\footnote{Here is a proof that uses the 3-bracket language and the fundamental identity,
\bea
&&[[B_{PQ},B_{MN},v^Q],v^P,\bullet] + [v^Q,[B_{PQ},B_{MN},v^P],\bullet] \cr
&&= [B_{PQ},B_{MN},[v^Q,v^P,\bullet]] - [v^Q,v^P,[B_{PQ},B_{MN},\bullet]] = 0
\eea}
\bea
\<(A_P,B_{MN}),v^P\> &=& 0
\eea
then we may write this result as
\ben
F_{MN} &=& \<H_{MNP},v^P\>\label{FH}
\een
where
\bea
H_{MNP} &=& 3\partial_{[M} B_{NP]} + \frac{3}{2} (A_{[M},B_{NP]}) + C_{MNP}\cr
\<C_{MNP},v^P\> &=& 0
\eea
provided that we impose the constraint
\bea
\L_v B_{MN} &=& 0
\eea
where 
\bea
\L_v B_{MN} &=& \<v^P,\partial_P B_{MN}\> + \<\partial_M v^P,B_{PN}\> + \<\partial_N v^P,B_{MP}\> 
\eea

We have no explicit expression for $C_{MNP}$, but we can derive its gauge variation by requiring that $H_{MNP}$ transforms homogeneosly. We compute the gauge variation as
\bea
\delta H_{MNP} &=& 3 \partial_M \delta B_{NP} + \frac{3}{2} (\delta A_M,B_{NP}) + \frac{3}{2} (A_M,\delta B_{NP}) + \delta C_{MNP}
\eea
We make the ansatz
\bea
\delta C_{MNP} &=& - (\Lambda,C_{MNP}) + \delta'C_{MNP}
\eea
for the gauge variation of $C_{MNP}$. We compute each term in turn,
\bea
3 \partial_M \delta B_{NP} &=& - 3 (\partial_M \Lambda,B_{NP}) - 3 (\Lambda,\partial_M B_{NP})\cr
\frac{3}{2} (\delta A_M,B_{NP}) &=& \frac{3}{2} (\partial_M \Lambda,B_{NP}) + \frac{3}{2} ([A_M,\Lambda],B_{NP})\cr
\frac{3}{2} (A_M,\delta B_{NP}) &=& 3 (A_M,\partial_N \Lambda_P) - \frac{3}{2} (A_M,(\Lambda,B_{NP}))
\eea
We then find the gauge variation,\footnote{Here we use the Jacobi identity
\bea
([A_M,\Lambda],B_{NP}) &=& (A_M,(\Lambda,B_{NP})) - (\Lambda,(A_M,B_{NP}))
\eea}
\bea
\delta H_{MNP} &=& - 3 (\Lambda,\partial_M B_{NP}) + \frac{3}{2} ([A_M,\Lambda],B_{NP}) - \frac{3}{2} (A_M,(\Lambda,B_{NP})) + (\Lambda,C_{MNP}) \cr
&=& - 3 (\Lambda,H_{MNP})
\eea
if we assume that 
\bea
\delta'C_{MNP} &=& \frac{3}{2} (\partial_M \Lambda,B_{NP}) - 3 (A_M,\partial_N \Lambda_P)
\eea
Let us summarize what we have found so far. Consistency with gauge symmetry of $A_M$ together with the minimal assumption that $A_M = \<B_{MN},v^N\>$ implies that the following conditions must be satisfied:
\begin{itemize}
\item The Lie derivatives along $v^M$ vanish, $\L_v B_{MN} = 0$ and $\L_v \Lambda_M = 0$.
\item We have the constraint $\<v^M,v^N\> = 0$.
\item We have the constraint $F_{MN} = \<H_{MNP},v^P\>$.
\item We have the 3-algebra Jacobi identities
\ben
\<(A,a),b\> - \<(A,b),a\> &=& [A,\<a,b\>]\label{Tre1}
\een
\ben
(A,(B,b)) - (B,(A,a)) &=& ([A,B],a)\label{Tre2}
\een
\ben
[[A,B],C] - [[A,C],B] &=& [A,[B,C]]\label{Tre3}
\een
for elements $a,b,...\in \A_3$ and $A,B,..\in \g$. 
\end{itemize}
The Jacobi identity (\ref{Tre3}) is the closure relation of the gauge algebra acting on a Lie algebra value field. The 3-algebra Jacobi identity (\ref{Tre2}) is the closure relation when acting on a 3-algebra valued field,
\bea
[\delta_{\Lambda'},\delta_{\Lambda}] \varphi &=& \delta_{[\Lambda,\Lambda']} \varphi
\eea
and the 3-algebra Jacobi identity (\ref{Tre1}) is the Leibniz rule for gauge transformations,
\bea
\delta_{\Lambda} \<\varphi_1,\varphi_2\> &=& \<\delta_{\Lambda}\varphi_1,\varphi_2\> + \<\varphi_1,\delta_{\Lambda}\varphi_2\>
\eea
This 3-algebra was first applied in a gauge theory in \cite{Gustavsson:2007vu}, \cite{Bagger:2007jr}, \cite{Bagger:2007vi}. In \cite{Bagger:2007vi} it was shown that the three Jacobi identities (\ref{Tre1}), (\ref{Tre2}), (\ref{Tre3}) are equivalent with the so-called fundamental identity, which is a generalized Jacobi identity satisfied by the 3-bracket $[a,b,c]$.

\section{Review of 6d $(2,0)$ classical field theory}
We begin by reviewing the 6d theory of \cite{Lambert:2010iw}, \cite{Chen:2013wya}, but where we drop the assumption that $v^M$ is a dynamical field that belongs to the tensor multiplet. We will instead assume that $v^M$ is a background vector field that does not belong to the tensor multiplet. As we will see below, without making this assumption, we can not extend the Poincare supersymmetry in \cite{Lambert:2010iw}, \cite{Chen:2013wya} to the full superconformal symmetry. Following \cite{Lambert:2010iw}, we assume that $\phi^A,\Psi,H_{MNP},v^M\in \A_3$ and $A_M \in \g$. Here $M = 0,1,2,3,4,5$ is a spacetime vector index and $A=1,2,3,4,5$ is an $SO(5)$ R-symmetry vector index. One may object to taking $v^M$ as a background field and at the same time assuming it is in $\A_3$ and hence transforming under gauge transformations as
\bea
\delta v^M &=& -(\Lambda,v^M)
\eea
where $\Lambda = \<\Lambda_N,v^N\>$. But this means that by the Abelian constraint $\<v^M,v^N\> = 0$, we actually have that $v^M$ transforms trivially under gauge transformations,
\bea
\delta v^M &=& 0
\eea
so there is no contradiction in assuming that $v^M$ is a background field. 

In 11d spinors have $32$ components. We break this down to $16$ components by imposing the 6d Weyl projections $\Gamma \Psi = \Psi$ and $\Gamma\epsilon = -\epsilon$ where $\Gamma = \Gamma^{012345}$. We use 11d gamma matrices that we denote as $\Gamma^M$ and $\h\Gamma^A$.\footnote{The hat notation is used just to distinugish say $\Gamma^1$ from $\h\Gamma^1$. We let $M=0,1,...,5$ be the spacetime vector index and $A=1,...,5$ be the R-symmetry index.} In 11d notation, these are satisfying the Clifford algebra
\bea
\{\Gamma^M,\Gamma^N\} &=& 2 \eta^{MN}\cr
\{\h\Gamma^A,\h\Gamma^B\} &=& 2 \delta^{AB}\cr
\{\Gamma^M,\h\Gamma^A\} &=& 0
\eea
where the spacetime metric is $\eta_{MN} = $diag$(-1,1,1,1,1,1)$. The $(2,0)$ Poincare supersymmetry variations on $\mb{R}^{1,5}$ read \cite{Lambert:2010iw}
\bea
\delta \phi^A &=& i \bar\epsilon\h\Gamma^A\Psi\cr
\delta \Psi &=& \frac{1}{12}\Gamma^{MNP}\epsilon H_{MNP} + \Gamma^M \h\Gamma^A \epsilon D_M \phi^A - \frac{1}{2} \Gamma^M \h\Gamma^{AB} \epsilon [\phi^A,\phi^B,v_M]\cr
\delta H_{MNP} &=& 3 i \bar\epsilon \Gamma_{NP} D_M \Psi + i \bar\epsilon \h\Gamma^A \Gamma_{MNPQ} [\Psi,v^Q,\phi^A]\cr
\delta A_M &=& i \bar\epsilon \Gamma_{MN} \<\Psi,v^N\>
\eea
We notice that the 3-bracket term in $\delta H_{MNP}$ vanishes when this is contracted by $v^P$. This follows from the Abelian constraint together with the Jacobi identity $\<(A,v^Q),v^P\> - \<(A,v^P),v^Q\> = [A,\<v^Q,v^P\>] = 0$ where we take $A = \<\Psi,\phi^A\>$. By using this, the variation of $H_{MNP}$ implies the following variation for $F_{MN} = \<H_{MNP},v^P\>$, 
\bea
\delta F_{MN} &=& 2 i \bar\epsilon \Gamma_{NP} \<D_M \Psi,v^P\> + i \bar\epsilon \Gamma_{MN} \<D_P \Psi,v^P\>
\eea
On the other hand, the variation of $A_M$ implies the variation $\delta F_{MN} = 2 D_M \delta A_N$, which is given by
\bea
\delta F_{MN} &=& 2 i \bar\epsilon \Gamma_{NP} \<D_M\Psi,v^P\> + 2 i \bar\epsilon \Gamma_{NP} \<\Psi,D_M v^P\>
\eea
Consistency requires the two expressions agree, which is the case only if
\bea
\L_v \(\bar\epsilon \Gamma_{MN} \psi\) &=& 0
\eea
Here this Lie derivative is given by
\bea
\L_v \(\bar\epsilon \Gamma_{MN} \psi\) &=& \<v^P,\bar\epsilon \Gamma_{MN} D_P \psi\> + \<D_M v^P,\bar\epsilon \Gamma_{PN} \psi\> + \<D_N v^P,\bar\epsilon\Gamma_{MP}\psi\>
\eea

To close these supersymmetry variations on-shell, we need the following constraints
\bea
F_{MN} &=& \<H_{MNP},v^P\>\cr
\<\L_v,{\mbox{any field in the tensor multiplet}}\> &=& 0\cr
\<v^M,v^N\> &=& 0
\eea
Then these supersymmetry variations close on-shell. In particular we note that the only constraint we need to impose on $v^M$ is the Abelian constraint. As was noted in \cite{Lambert:2010iw}, there are two ways we can solve the second constraint. If the 3-algebra is nontrivial, that is, if the gauge group is non-Abelian, then we need the Lie derivative of all the fields to vanish. But if the 3-algebra is trivial so that all 3-brackets vanish and the gauge group is Abelian, then we do not need any Lie derivatives to vanish and we get back the usual Abelian $(2,0)$ tensor multiplet. So the Abelian case is included in this formulation, which is a nice property to have for a theory that should be a generalization of the Abelian theory.

We have the following fermionic equation of motion,
\bea
\Gamma^M D_M \Psi + \Gamma^M \h\Gamma^A [\Psi,v_M,\phi^A] &=& 0
\eea
that can be derived from the Lagrangian 
\ben
\L_{\Psi} &=& \frac{i}{2} \bar\Psi \cdot \Gamma^M D_M \Psi + \frac{i}{2} \bar\Psi \cdot \Gamma^M \h\Gamma^A [\Psi,v_M,\phi^A]\label{Psi}
\een

\subsection{Reducing down to $(1,0)$ supersymmetry} 
To reduce down to $(1,0)$ supersymmetry we let the supersymmetry parameter be subject to the Weyl projection \cite{Chen:2013wya}
\bea
\h\Gamma \epsilon &=& - \epsilon
\eea
where $\h\Gamma = \h\Gamma^{1234}$. If we assume a sign convention such that $\h\Gamma^5 = \Gamma \h\Gamma$, then 
\ben
\h\Gamma^5 \epsilon &=& \epsilon\label{hW}
\een
The $(2,0)$ tensor multiplet separates into one $(1,0)$ tensor multiplet and one $(1,0)$ hypermultiplet. For the tensor multiplet fermions $\psi$, we have
\bea
\h\Gamma \psi &=& - \psi\cr
\h\Gamma^5 \psi &=& - \psi
\eea
and for the hypermultiplet fermions $\chi$, we have
\bea
\h\Gamma \chi &=& \chi\cr
\h\Gamma^5 \chi &=& \chi
\eea
A consequence of this Weyl projection is the $SO(5)$ R symmetry is broken to $SU(2)_R \times SU(2)_F$. We note that $\h\Gamma^{ij}$ acting on $\psi$ belongs to the same $SU(2)$ as when it acts on $\epsilon$ since both $\psi$ and $\epsilon$ have the same $\h\Gamma$ chirality, so this is in $SU(2)_R$, the R-symmetry. Whereas $\h\Gamma^{ij}$ when acting on $\chi$ is an element in the flavor symmetry $SU(2)_F$. Thus $SO(5)_R$ R-symmetry is reduced to $SU(2)_R$ R-symmetry, and in addition there is a global $SU(2)_F$ symmetry. We split the index $A = \{i,5\}$ where $i = 1,2,3,4$ and we will use the notation $\sigma = \phi^5$. The supersymmetry variations are, for the tensor multiplet,
\bea
\delta \sigma &=& - i \bar\epsilon \psi\cr
\delta \psi &=& \frac{1}{12} \Gamma^{MNP} \epsilon H_{MNP} + \Gamma^M \epsilon D_M \sigma - \frac{1}{2} \Gamma^M \h\Gamma^{ij} \epsilon [\phi^i,\phi^j,v_M]\cr
\delta H_{MNP} &=& 3 i \bar\epsilon \Gamma_{NP} D_M \psi - i \bar\epsilon \Gamma_{MNPQ} [\psi,v^Q,\sigma]\cr
\delta A_M &=& i \bar\epsilon \Gamma_{MN} \<\psi,v^N\>
\eea
and for the hypermultiplet,
\bea
\delta \phi^i &=& i \bar\epsilon \h\Gamma^i \chi\cr
\delta \chi &=& \Gamma^M \h\Gamma^i \epsilon D_M \phi^i - \Gamma^M \h\Gamma^i \epsilon [\phi^i,\sigma,v_M]
\eea
We get the fermionic equations of motion
\bea
\Gamma^M D_M \psi + \Gamma^M \h\Gamma^i [\chi,v_M,\phi^i] - \Gamma^M [\psi,v_M,\sigma] &=& 0\cr
\Gamma^M D_M \chi + \Gamma^M \h\Gamma^i [\psi,v_M,\phi^i] + \Gamma^M [\chi,v_M,\sigma] &=& 0
\eea
which can be integrated up to the Lagrangian
\bea
\L_{\Psi} &=& \frac{i}{2} \bar\chi \cdot \Gamma^M D_M \chi + \frac{i}{2} \bar\chi\cdot\Gamma^M[\chi,v_M,\sigma]\cr
&+& \frac{i}{2} \bar\psi\cdot \Gamma^M D_M\psi - \frac{i}{2} \bar\psi\cdot \Gamma^M[\psi,v_M,\sigma]\cr
&+& i \bar\psi \cdot \Gamma^M \h\Gamma^i [\chi,v_M,\phi^i]
\eea
The last term is a couping term between tensor and hypermultiplet fermions, which makes it hard to separate this Lagrangian into one Lagrangian for the tensor multiplet and another for the hypermultiplet. The best we can do is to treat them together, although the equations of motion and the superrsymmetry variations can be treated separately for the tensor and the hypermultiplets.

\section{Superconformal symmetry}
We now ask ourselves if the theories of \cite{Lambert:2010iw}, \cite{Chen:2013wya} are superconformal at the classical level. Our starting point is to assume that we have a supersymmetry parameter that is a conformal Killing spinor $\epsilon_f$ satisfying
\ben
D_M \epsilon_f &=& \Gamma_M \eta\label{CKE}
\een
Since we do not have a formulation in terms of a 2-form gauge potential $B_{MN}$, it is not apriori clear how we shall generalize the supersymmetry variation for $H_{MNP}$ in \cite{Lambert:2010iw}, \cite{Chen:2013wya} to the case when $\epsilon_f$ is not a constant spinor. We will make the following ansatz for the $(1,0)$ tensor multiplet,
\ben
\delta H_{MNP} &=& 3 i \bar\epsilon_f \Gamma_{[NP} D_{M]}\psi - 3 i \bar\eta\Gamma_{MNP} \psi - i \bar\epsilon_f\Gamma_{MNPQ} [\psi,v^Q,\sigma]\cr
\delta A_M &=& i \bar\epsilon_f\Gamma_{MN} \<\psi,v^N\>\cr
\delta \sigma &=& - i \bar\epsilon_f \psi\cr
\delta \psi &=& \frac{1}{12} \Gamma^{MNP} \epsilon_f H_{MNP} + \Gamma^M \epsilon_f D_M \sigma + 4 \eta \sigma\label{CFT}
\een
Here a coupling term to the hypermultiplet has been put to zero for simplicity. This ansatz for the supersymmetry variations is consistent with the relation\footnote{To see this, we use the gamma matrix identity
\bea
3 \Gamma_{MNP} &=& 2\Gamma_{[M} \Gamma_{N]P} + \Gamma_P \Gamma_{MN} 
\eea}
\ben
\delta F_{MN} &=& \<\delta H_{MNP},v^P\> + i \<\L_v,\bar\epsilon \Gamma_{MN} \psi\>\label{FHvariation}
\een
where 
\bea
\<\L_v,\bar\epsilon \Gamma_{MN} \psi\> &:=& \<v^P,D_P\(\bar\epsilon\Gamma_{MN}\psi\)\> + \<D_M v^P,\bar\epsilon\Gamma_{PN}\psi\> + \<D_N v^P,\bar\epsilon\Gamma_{MP}\psi\>
\eea
Thus, in order for the ansatz to be compatible with the constraint (\ref{FH}), we need to impose the constraints
\bea
\<\L_v,\psi\> &=& 0\cr
\L_v \epsilon_f &=& 0
\eea
where
\bea
\<\L_v,\psi\> &=& \<v^P,D_P\psi\> + \frac{1}{4} \<D_M v_N,\Gamma^{MN}\psi\>\cr
\L_v \epsilon_f &=& v^P D_P \epsilon_f + \frac{1}{4} D_M v_N \Gamma^{MN} \epsilon_f
\eea
In flat $\mb{R}^{1,5}$ we can solve the conformal Killing spinor equation (\ref{CKE}). The solution is given by
\bea
\epsilon_f &=& \epsilon + \Gamma_M \eta x^M
\eea
where $\epsilon$ gives the Poincare supersymmetry and $\eta$ gives the special conformal supersymmetry. In this paper, we will assume that $\epsilon_f$ and all associated supersymmetry parameters (such as $\epsilon$ and $\eta$) are commuting spinors. This means that the variation $\delta_{\epsilon_f}$ is anticommuting. When closing the supersymmetry variations, it is then enough to just compute $\delta_{\epsilon_f}^2$, and this will provide for us the most general supersymmetry closure relations. Let us spend some lines on explaining this point in some detail here. If we consider the perhaps more familiar situation with say two anticommuting Poincare supersymmetry parameters $\epsilon$ and $\epsilon'$, then we will compute the commutator $[\delta_{\epsilon},\delta_{\epsilon'}]$ in order to check the most general closure of these supersymmetry variations. If we instead have commuting supersymmetry parameters, we should correspondingly compute the anticommutator $\{\delta_{\epsilon},\delta_{\epsilon'}\}$ between two different supersymmetry variations, and not just the square $\delta_{\epsilon}^2$ of some supersymmetry variation. But as long as the supersymmetry variations are linear in the supersymmetry parameter, we may indeed equally well just compute the squares of supersymmetry variations. The reason why this will be sufficient, is because of linearity
\bea
\delta_{\epsilon+\epsilon'} &=& \delta_{\epsilon} + \delta_{\epsilon'}
\eea
and because we have the identity
\ben
\{\delta_{\epsilon},\delta_{\epsilon'}\} &=& \delta_{\epsilon+\epsilon'}^2 - \delta_{\epsilon}^2 - \delta_{\epsilon'}^2\label{g}
\een
where on the right-hand side only appears perfect squares. So by just computing $\delta_{\epsilon}^2$ for a general parameter $\epsilon$, we can extract the most general anticommutator closure relations by using (\ref{g}). 

Inspired by the closure computation for 3d superconformal transformations in \cite{Chen:2012kk}, we will perform the same type of closure computation here for our 6d $(1,0)$ theory. We will find that the above supersymmetry transformations close into the generators of the conformal group up to a gauge transformation,
\bea
\delta A_M &=& D_M \Lambda\cr
\delta \sigma &=& - (\Lambda,\sigma)\cr
\delta \psi &=& - (\Lambda,\psi)
\eea 
But the closure relation is not very clear to us for a generic 6d manifold, so in the end we specialize to flat $\mb{R}^{1,5}$ where interpretations in terms of generators of the superconformal algebra is clear. 

Let us introduce the quantity
\ben
S^M &:=& \bar\epsilon_f \Gamma^M \epsilon_f\label{CKV}
\een
By using (\ref{CKE}) we get the derivatives 
\bea
D_M S_N &=& 2 \bar\epsilon_f \eta g_{MN} - 2\bar\epsilon_f \Gamma_{MN} \eta\cr
D^M S_M &=& 12 \bar\epsilon_f \eta
\eea

In flat $\mb{R}^{1,5}$ we have the expansion
\bea
S^M \P_M &=& \bar\epsilon \Gamma^M \epsilon \P_M + 2 \bar\epsilon \eta \D + \bar\epsilon \Gamma^{MN} \eta \L_{MN} - \bar\eta \Gamma^M \eta \K_M 
\eea
where we define 
\bea
\P_M &=& - i \partial_M\cr
\D &=& - i x^M \partial_M\cr
\L_{MN} &=& i (x_M \partial_N - x_N \partial_M)\cr
\K_M &=& i (-2x_M x^N \partial_N + |x|^2 \partial_M)
\eea
As we will show below, we find the following closure relation when we act on any of the fields in the tensor multiplet in flat space,
\bea
\delta^2 &=& \bar\epsilon \Gamma^M \epsilon P_M + 2 \bar\epsilon \eta D + \bar\epsilon \Gamma^{MN} \eta L_{MN} - \bar\eta \Gamma^M \eta K_M - 2 \bar\epsilon\h\Gamma^{ij}\eta S^{ij}
\eea
where the generators of the conformal group are given by \cite{Qualls:2015qjb}
\bea
P_M &=& \P_M\cr
D &=& \D - i\Delta\cr
L_{MN} &=& \L_{MN} + S_{MN}\cr
K_M &=& \K_M - 2 i \Delta x_M - S_{MN} x^N
\eea
Here $\Delta$ is the scaling dimension of the field, and $S_{MN}$ is the spin part of the Lorentz rotation, which we normalize such that 
\bea
S_{MN} &=& \frac{i}{2} \Gamma_{MN}\cr
S_{MN}^{PQ} &=& 2i \delta_{MN}^{PQ}
\eea
when acting on a spinor and a vector field respectively. The last generator $S^{ij}$ is a generator for the R-symmetry $SU(2)_R$ and it acts nontrivially only on the tensor muliplet fermions and the hypermultiplet scalars. 

We will view $v^M$ as a background field, perhaps a bit similar to the background metric tensor field $g_{MN}$. In supersymmetric field theory, as opposed to supergravity, we normally do not include $g_{MN}$ in the supermultiplet and demand its supersymmetry variation is vanishing,
\bea
\delta g_{MN} &=& 0
\eea
But let us assume that we do this anyway, just to see where this can lead us. Then we shall require supersymmetric closure on the metric tensor,
\bea
\delta^2 g_{MN} &=& - i \L_S g_{MN}
\eea
We then reach the conclusion that $S^M$ shall be a Killing vector in order for supersymmetry closure,
\ben
\L_S g_{MN} &=& 0\label{P}
\een
which shows that extension to superconformal symmetry where $S^M$ is a conformal Killing vector,
\ben
\L_S g_{MN} &=& \frac{1}{3} g_{MN} D^P S_P\label{CKE1}
\een
is not possible. We note that (\ref{CKE}) implies (\ref{CKE1}) but not necessarily (\ref{P}) which is too restrictive. Similarly, if we include $v^M$ in the supersymmetry variations, we can not generalize Poincare supersymmery to the full superconformal symmetry. The closure relation we would need to require is given by
\bea
\delta^2 v_M &=& - i \L_S v_M
\eea
and what we get from assuming the supersymmetry variation
\bea
\delta v_M &=& 0
\eea
is $\delta^2 v_M = 0$ and so we would have to require that the Lie derivative vanishes,
\bea
S^N D_N v_M + (D_M S^N) v_N &=& 0
\eea
If we expand this out on $\mb{R}^{1,5}$, we get
\bea
\bar\epsilon \Gamma^N \epsilon D_N v_M + 2\bar\epsilon \eta \( -i x^N D_N + 1\) v_M + ... &=& 0
\eea
Since $\bar\epsilon \Gamma^N \epsilon$, $\bar\epsilon \eta$, ... are all independent, each term must vanish separately. The vanishing of the first term leads to the constraint
\bea
D_N v_M &=& 0
\eea
which is a constraint that was found in \cite{Lambert:2010iw}. The vanishing of the second term leads to $\eta = 0$. That is, only the Poincare supersymmetry can be realized. 

Since we want the tensor multiplet to be superconformal, we are led to assume that $v^M$ shall be a background field that does not belong to the tensor multiplet. 

We will now verify that we have the appropriate closure relations when we act on anyone of the fields in the tensor multiplet. But this check does not include the vector field $v^M$, which therefore requires a separate treatment.

\subsubsection*{Closure on the scalar field}
For the scalar field $\sigma$ we get
\bea
\delta^2 \sigma &=& - i S^M D_M \sigma - \frac{i}{3} (D^M S_M) \sigma
\eea
We can also write this in the form
\bea
\delta^2 \sigma &=& - i S^M \partial_M \sigma - \frac{i}{3} (D^M S_M) \sigma - (\Lambda,\sigma)
\eea
where
\bea
\Lambda &=& i S^M \(A_M + \<v_M,\sigma\>\)
\eea
Expanding this out, we get
\bea
\delta^2 \sigma &=& \bar\epsilon\Gamma^M\epsilon \P_M \sigma\cr
&& + 2 \bar\epsilon\eta \(\D - 2i\) \sigma\cr
&& + \bar\epsilon\Gamma^{PQ}\eta \L_{PQ}\sigma\cr
&& - \bar\eta \Gamma^P\eta \(\K_P - 4i x_P\) \sigma
\eea
Thus we get closure, and the second line is telling us that the scaling dimension of $\sigma$ is $\Delta = 2$.

\subsubsection*{Closure on the gauge field}
For the gauge field $A_M$ we get 
\bea
\delta^2 A_M &=& - i S^R \<H_{RMN},v^N\> + D_M \Lambda'\cr
&& + \frac{i}{2} S^R \<H^-_{RMN},v^N\>\cr
&& + i S_M \<\L_v,\sigma\>\cr
&& + i \<\sigma,\L_v S_M\>
\eea
where
\bea
\<\L_v,\sigma\> &:=& \<D_N \sigma,v^N\>\cr
\L_v S_M &:=& v^N D_N S_M + (D_M v^N) S_N
\eea
and 
\bea
\Lambda' &=& i S^M \<v_M,\sigma\>
\eea  
Selfduality of $H^+_{MNP}$ is connected with Weyl projection of $\epsilon_f$, such that 
\bea
\Gamma^{MNP} \epsilon H^{-}_{MNP} &=& 0
\eea
With our convention, 
\bea
H^{\pm}_{MNP} &=& \frac{1}{2} \(H_{MNP} \pm \frac{1}{6} \epsilon_{MNP}{}^{RST} H_{RST}\)
\eea
To get the closure relation, we have also noted that 
\bea
2\bar\epsilon_f \Gamma_{MN} \eta &=& - D_M S_N + D_N S_M
\eea
We thus we find the closure relation
\bea
\delta^2 A_M &=& - i S^R F_{RM} + D_M\Lambda'
\eea
if we impose the constraints
\bea
H^-_{MNP} &=& 0\cr
\<\L_v,\sigma\> &=& 0\cr
\L_v S_M &=& 0
\eea
We can write this as
\bea
\delta^2 A_M &=& - i \L_S A_M + D_M\Lambda
\eea
where the gauge parameter is the same as before when we closed supersymmetry on $\sigma$,
\bea
\Lambda &=& i S^M \(A_M + \<v_M,\sigma\>\)
\eea
and the Lie derivative is given by
\bea
\L_S A_M &=& S^N \partial_N A_M + (\partial_M S^N) A_N
\eea
If we expand out this Lie derivative explicitly, we get
\bea
-i \L_S A_M &=& \bar\epsilon \Gamma^P \epsilon \P_P A_M\cr
&& + 2 \bar\epsilon\eta \(\D-i\) A_M\cr
&& + \bar\epsilon \Gamma^{PQ} \eta \(\L_{PQ} \delta_M^N + (S_{PQ})_M{}_N\) A_N\cr
&& - \bar\eta \Gamma^P \eta \(\K_P \delta_M^N - 2(S_{PQ})_M{}^N - 2 i x_P \delta_M^N\) A_N
\eea
Thus we closure and the second line is telling us that the scaling dimension of $A_M$ is $\Delta = 1$.

\subsubsection*{Closure on the fermions}
For the fermions $\psi$ we get 
\bea
\delta^2\psi &=& i \(\frac{1}{4} \Gamma^{MNP} \epsilon_f \bar\epsilon_f \Gamma_{NP} - \Gamma^M \epsilon_f \bar\epsilon_f \) D_M \psi\cr
&& + i \(- \frac{1}{4} \Gamma^{MNP} \epsilon_f \bar\eta \Gamma_{MNP} +  \Gamma^M \epsilon_f \bar\eta \Gamma_M - 4 \eta \bar\epsilon_f\) \psi\cr
&& + i \(- \frac{1}{12} \Gamma^{MNP} \epsilon_f\bar\epsilon_f \Gamma_{MNPQ} +  \Gamma^M \epsilon_f \bar\epsilon_f \Gamma_{MQ}\) [\psi,v^Q,\sigma]
\eea
We have the following Fierz identities,
\bea
\epsilon_f \bar\epsilon_f &=& \frac{1}{4}(1-\Gamma)(1-\h\Gamma) \(c^M \Gamma_M + c^{MNP,ij} \Gamma_{MNP}\h\Gamma^{ij}\)\cr
\epsilon_f \bar\eta &=& \frac{1}{4}(1-\Gamma)(1-\h\Gamma) \(c + c^{MN} \Gamma_{MN} + c^{ij} \h\Gamma^{ij} + c^{MN,ij} \Gamma_{MN} \h\Gamma^{ij}\)\cr
\eta\bar\epsilon_f &=& \frac{1}{4}(1+\Gamma)(1-\h\Gamma) \(-c + c^{MN} \Gamma_{MN} + c^{ij} \h\Gamma^{ij} - c^{MN,ij} \Gamma_{MN}\h\Gamma^{ij}\)
\eea
where
\bea
c^M &=& \frac{1}{8} \bar\epsilon_f \Gamma^M \epsilon_f\cr
c^{MNP,ij} &=& \frac{1}{96} \bar\epsilon_f\Gamma^{RST}\h\Gamma^{ij}\epsilon_f\cr
c &=& - \frac{1}{8} \bar\epsilon_f \eta\cr
c^{ij}_+ &=& - \frac{1}{32} \bar\epsilon_f \h\Gamma^{ij} \eta\cr
c^{MN} &=& - \frac{1}{16} \bar\epsilon_f \Gamma^{MN} \eta\cr
c^{MN,ij} &=& - \frac{1}{64} \bar\epsilon_f\Gamma^{MN}\h\Gamma^{ij}\eta
\eea
We have indicated with a subscript $+$ the fact that $c^{ij}_+$ is selfdual. We separate $\delta^2 \psi$ into three parts,
\bea
(\delta^2 \psi)_{abel} &=& - 8 i c^M D_M \psi + 2 i c^Q \Gamma_Q \Gamma^M D_M \psi\cr
(\delta^2 \psi)_{curv} &=& \(40 i c - 8 i \Gamma_{MN} c^{MN} + 32 i c^{ij} \h\Gamma^{ij}\) \psi\cr
(\delta^2 \psi)_{comm} &=& - \( 8 i c^M + 2 i c^R \Gamma_R \Gamma^M\) [\psi,v_M,\sigma]
\eea
Let us first rewrite 
\bea
(\delta^2 \psi)_{abel} &=& - 8 i c^M \partial_M \psi + 2 i c^Q \Gamma_Q \Gamma^M D_M \psi - (\Lambda'',\psi)
\eea
where
\bea
\Lambda'' &=& 8 i c^M A_M
\eea
Then we expand out the sum of the three terms in components to get
\bea
\delta^2 \psi &=& \bar\epsilon \Gamma^P \epsilon \P_P \psi\cr
&& + 2 \bar\epsilon\eta \(\D-\frac{5i}{2}\) \psi\cr
&& + \bar\epsilon \Gamma^{PQ} \eta \(\L_{PQ} + S_{PQ}\) \psi\cr
&& - \bar\eta \Gamma^P \eta \(\K_P - 2 S_{PQ}x^Q - 5 i x_P\) \psi\cr
&& - 2\bar\epsilon \h\Gamma^{ij} \eta S^{ij} \psi\cr
&& - (\Lambda,\psi)\cr
&& + 2 i c^Q \Gamma_Q \(\Gamma^M D_M \psi - \Gamma^M [\psi,v_M,\sigma]\)
\eea 
where the gauge parameter is 
\bea
\Lambda &=& i S^M \(A_M + \<v_M,\sigma\>\)
\eea
Thus we have closure and the second line is telling us that the scaling dimension of $\psi$ is $\Delta = 5/2$.

\subsubsection*{Closure on the tensor field}
For the tensor field $H_{MNP}$ we get
\ben
(\delta^2 H_{MNP})_{abel} &=& \frac{i}{4} D_M \(\bar\epsilon_f \Gamma_{NP} \Gamma^{RST} \epsilon_f H_{RST}\) \label{h1}\\
&& + 3 i \bar\epsilon_f \Gamma_{NP}\Gamma^R \epsilon_f D_M D_R \sigma\label{h2}\\
&& + 12 i \bar\epsilon_f \Gamma_{NP} \eta D_M \sigma\label{h3}\\
(\delta^2 H_{MNP})_{curv} &=& 3 i \bar\epsilon_f \(\Gamma_{NP}\Gamma^R\Gamma_M - \Gamma^R\Gamma_{MNP}\) \eta D_R\sigma\label{h5}\\
(\delta^2 H_{MNP})_{comm} &=& - \frac{i}{12} \bar\epsilon_f \Gamma_{MNPQ} \Gamma^{RST} \epsilon_f [H_{RST},v^Q,\sigma]\label{h6}\\
&& - i \bar\epsilon_f \Gamma_{MNPQ} \Gamma^R \epsilon_f [D_R \sigma,v^Q,\sigma]\label{h7}\\
&& - 3 [\bar\epsilon_f \Gamma_{NP} \psi,\bar\epsilon_f \Gamma_{MQ} \psi,v^Q]\label{h8}
\een
We use
\bea
\Gamma_{NP} \Gamma^R \Gamma_M &=& - \frac{2}{3} \Gamma_{MNP}\Gamma^R + \frac{1}{3} \Gamma^R \Gamma_{MNP}
\eea
to rewrite
\bea
(\delta^2 H_{MNP})_{curv} &=& - 12 i \bar\epsilon_f \Gamma_{MN} \eta D_P \sigma
\eea
which then cancels against (\ref{h3}). We next look at the commutator terms. We have
\bea
\bar\epsilon_f \Gamma_{MNPQ} \Gamma^{RST} \epsilon_f &=& 12 \bar\epsilon_f\(\delta_{MNP}^{RST} \Gamma_Q - 3 \delta_{Q[MN}^{RST} \Gamma_{P]}\)\epsilon_f
\eea
This then enables us to rewrite (\ref{h6}) as
\ben
- (\Lambda'',H_{MNP}) + 3 i S_{[P} [H_{MN]Q},v^Q,\sigma]\label{h6prime}
\een
where
\bea
\Lambda'' &=& i S^M \<v_M,\sigma\>
\eea
Let us return to the abelian type of terms,
\bea
(\delta^2 H_{MNP})_{abel} &=& - i \(S^Q D_Q H_{MNP} + 3 (D_M S^Q) H_{NPQ}\)\cr
&& - 4 i S^Q D_{[M} H_{NPQ]}\cr
&& - 3 i S_{[P} (F_{MN]},\sigma)
\eea
The last term cancels against (\ref{h6prime}) if we assume that
\bea
F_{MN} &=& \<H_{MNP},v^P\>
\eea
The remaining terms in $(\delta^2 H_{MNP})_{comm}$ cancel by modifying the Bianchi identity \cite{Lambert:2010iw}, \cite{Chen:2013wya}. Let us now summarize the closure relation that we have got,
\bea
\delta^2 H_{MNP} &=& - i \(S^Q D_Q H_{MNP} + 3 (D_M S^Q) H_{NPQ}\) - (\Lambda'',H_{MNP})
\eea
We can now extract the Lie derivative,
\bea
\delta^2 H_{MNP} &=& - i \L_S H_{MNP} - (\Lambda,H_{MNP})
\eea
where
\bea
\L_S H_{MNP} &=& S^Q \nabla_Q H_{MNP} + 3 (\nabla_M S^Q) H_{NPQ}\cr
&=& S^Q \partial_Q H_{MNP} + 3 (\partial_M S^Q) H_{NPQ}
\eea
where we have used the fact that $D_M S^Q = \nabla_M S^Q$ since $S^Q$ is a gauge singlet, so only the Christoffel symbol in $D_M$ acts on it nontrivially. We thus decompose $D_M = \nabla_M + A_M$.  

If we finally expand out the Lie derivative, we can see the conformal generators and the scaling dimension explicitly,
\bea
- i \L_S H_{MNP} &=& \bar\epsilon \Gamma^Q \epsilon \P_Q H_{MNP}\cr
&& + 2 \bar\epsilon\eta \(\D-3i\) H_{MNP}\cr
&& + \bar\epsilon \Gamma^{PQ} \eta \(\L_{PQ} + S_{PQ}\) H_{MNP}\cr
&& - \bar\eta \Gamma^P \eta \(\K_P - 2 S_{PQ}x^Q - 6 i x_P\) H_{MNP}
\eea
Thus we have closure and the second line is telling us that the scaling dimension of $H_{MNP}$ is $\Delta = 3$.

This finishes our on-shell closure computation for the $(1,0)$ tensor multiplet. It shows that the vector field $v^M$ shall be constrained by two relations,
\ben
\<v^M,v^N\> &=&0\label{Abelianconstraint}\\
\L_v \epsilon_f &=& 0\label{VanishingKilling}
\een
in order to have superconformal symmetry. 

Let us now return to our vector field $v^M$. Since scaling dimension for $H_{MNP}$ is $\Delta=3$ and for $F_{MN}$ it is $\Delta=2$, in order for the constraint $F_{MN} = \<H_{MNP},v^P\>$ to transform covariantly under conformal transformations, we shall assign $v^M$ scaling dimension $\Delta = -1$ and spin one. 

\section{Off-shell supersymmetry for the 6d hypermultiplet}
Off-shell supersymmetry for 10d SYM was first found in \cite{Berkovits:1993hx} and later this was used in \cite{Pestun:2007rz} for localization computations. There are $9$ off-shell components for the gauge potential after using the Gauss law constraint, $16$ off-shell components for the spinor. So we need $7$ auxiliary bosonic field components in order to make the number of off-shell bosonic degrees of freedom match with the number of off-shell fermionic degrees of freedom. If we dimensionally reduce to 5d we expect to find an off-shell tensor multiplet (with $3$ auxiliary fields) coupled to an off-shell hypermultiplet (with $4$ off-shell fields). But instead of following such an approach, there was a direct construction of an off-shell 5d tensor multiplet coupled to an off-shell 5d hypermultiplet appearing in \cite{Hosomichi:2012ek}.

One may ask whether there exists a 6d uplift of this off-shell supersymmetry. For the 6d hypermultiplet we can have off-shell supersymmetry since we can get this multiplet by reducing 10d $\N=1$ SYM down to $\N=(1,1)$ SYM in 6d \cite{Minahan:2015jta}. Reducing supersymmetry further down to $\N=(1,0)$, we get a 6d vector multiplet coupled to a 6d hypermultiplet. So in 6d we can have an off-shell hypermultiplet as well as an off-shell vector multiplet, but by such a construction, none of these multiplets will be conformal. The 6d vector multiplet can not be conformal. But as we will show, the 6d hypermultiplet can become superconformal. This is intuitively clear because not only can we get the $\N=(1,0)$ hypermultiplet from reducing the nonconformal $\N=(1,1)$ 6d vector multiplet down to $\N=(1,0)$, but also from the reducing the superconformal 6d $\N=(2,0)$ tensor multiplet down to $\N = (1,0)$.

\subsection{Abelian gauge group and Poincare supersymmetry}
We make the following ansatz for the off-shell supersymmetry variations,
\bea
\delta \phi^i &=& i \bar\epsilon\h\Gamma^i\chi\cr
\delta \chi &=& \Gamma^M \h\Gamma^i \epsilon \partial_M \phi^i + \sum_{\alpha=1}^n \nu^{\alpha} F^{\alpha}\cr
\delta F^{\alpha} &=& - i \bar\nu^{\alpha} \Gamma^M \partial_M \chi 
\eea
From now, we will suppress the summation symbol when $\alpha$ is contracted, and always assume Einstein summation convention. For closure on $\phi^i$, we compute
\bea
\delta^2 \phi^i &=& - i\bar\epsilon\h\Gamma^M\epsilon \partial_M\phi^i + i\bar\epsilon\h\Gamma^i\nu^{\alpha}F^{\alpha}
\eea
Thus closure on $\phi^i$ requires that 
\ben
\bar\epsilon\h\Gamma^i\nu^{\alpha} &=& 0\label{B1}
\een
Next,
\bea
\delta^2 F^{\alpha} &=& - i \bar\nu^{\alpha} \Gamma^M \nu^{\beta} \partial_M F^{\beta}
\eea
Thus closure on $F^{\alpha}$ requires in addition that 
\ben
\bar\nu^{\alpha} \Gamma^M \nu^{\beta} &=& \delta^{\alpha\beta} \bar\epsilon \Gamma^M \epsilon\label{B2}
\een
Finally, for the fermion, we get
\bea
\delta^2 \chi &=& i \(\Gamma^M \h\Gamma^i \epsilon\bar\epsilon \h\Gamma^i - \nu^{\alpha}\bar\nu^{\alpha} \Gamma^M\) \partial_M \chi
\eea
Using Fierz identities and assuming (\ref{B3}) holds, we get
\bea
\delta^2 \chi &=& i \frac{S^Q}{8} \(- 4 \Gamma^M \Gamma^Q - n \Gamma^Q \Gamma^M\) \partial_M \chi
\eea
and thus by taking $n = 4$ we get off-shell closure
\bea
\delta^2 \chi &=& - i S^Q \partial_Q \chi
\eea
but we also need to satisfy the constraints
\ben
\bar\nu^{\alpha} \Gamma^{MNP} \h\Gamma^{ij} \nu^{\alpha} &=& 0\label{B3}
\een

We have found the constraints 
\ben
\bar\epsilon\h\Gamma^i\nu^{\alpha} &=& 0\label{B1}\\
\bar\nu^{\alpha} \Gamma^M \nu^{\beta} &=& \delta^{\alpha\beta} \bar\epsilon \Gamma^M \epsilon\label{B2}\\
\bar\nu^{\alpha} \Gamma^{MNP} \h\Gamma^{ij} \nu^{\alpha} &=& 0\label{B3}
\een
Since we need $n=4$, we can try to make the ansatz
\bea
\nu^i &=& \h\Gamma^i \nu
\eea
in order to solve them. The index $\alpha$ has now been identified as the index $i$. Plugging in this ansatz, we automatically solve (\ref{B3}) since $\h\Gamma^i \h\Gamma^{kl} \h\Gamma^i = 0$, and the constraints reduce to
\bea
\bar\epsilon\nu &=& 0\cr
\bar\epsilon\h\Gamma^{ij}\nu &=& 0\cr
\bar\epsilon\Gamma^M \epsilon &=& \bar\nu\Gamma^M\nu
\eea
To understand how the last constraint appears, we compute
\bea
\bar\nu^i \Gamma^M \nu^j &=& - \bar\nu \h\Gamma^i \Gamma^M \h\Gamma^j \nu\cr
&=& \delta^{ij} \bar\nu \Gamma^M \nu
\eea
Up to a minus sign that depends on the convention we use for the charge conjugation marix, these are now the same constraints\footnote{More precisely, they are equivalent to, but we have not yet written them in the same form as they appeared in \cite{Hosomichi:2012ek}. For now, let us just note that we have $1+3$ constraints $\bar\epsilon\nu = 0$, $\bar\epsilon\h\Gamma^{ij}\nu = 0$
while \cite{Hosomichi:2012ek} has $2\times 2$ constraints of the form $\bar\epsilon_I \nu_J = 0$. We present all details how to convert between the two languages below.}, as those that appeared in 5d in \cite{Hosomichi:2012ek} and so these constraints can be solved as it was shown there. 

\subsection{Abelian gauge group and superconformal symmetry}
We make the ansatz
\bea
\delta \phi^i &=& i \bar\epsilon_f \h\Gamma^i\chi\cr
\delta \chi &=& \Gamma^M \h\Gamma^i \epsilon_f \partial_M \phi^i + \nu^{\alpha}_f F^{\alpha} - 4 \h\Gamma^i \eta \phi^i\cr
\delta F^{\alpha} &=& - i\bar\nu^{\alpha} \Gamma^M D_M \chi 
\eea
where we assume that 
\bea
D_M \epsilon_f &=& \Gamma_M \eta\cr
D_M \nu_f &=& \Gamma_M \rho
\eea
In flat space we solve these equations as
\bea
\epsilon_f &=& \epsilon + \Gamma_M \eta x^M\cr
\nu_f &=& \nu + \Gamma_M \rho x^M
\eea
We now proceed to check the closure relations. For the scalars we get 
\bea
\delta^2 \phi^i &=& - i S^M \partial_M \phi^i - 4 i\bar\epsilon_f \eta \phi^i - 4i \bar\epsilon_f \h\Gamma^{ij} \eta \phi^j
\eea
provided that we satisfy the constraint
\ben
\bar\epsilon_f \h\Gamma^i \nu_f^{\alpha} &=& 0\label{Co1}
\een
In flat space the closure relation becomes
\bea
\delta^2 \phi^i &=& \bar\epsilon\Gamma^M\epsilon \P_M \phi^i\cr
&& + 2 \bar\epsilon\eta \(\D - 2i\) \phi^i\cr
&& + \bar\epsilon\Gamma^{PQ}\eta \L_{PQ} \phi^i\cr
&& - \bar\eta \Gamma^P\eta \(\K_P - 4i x_P\) \phi^i\cr
&& - 2\bar\epsilon\h\Gamma^{kl}\eta (S^{kl})^{ij} \phi^j
\eea
Thus we have closure and the second line is telling us that the scaling dimension of $\phi^i$ is $\Delta=2$. We note that the last line is the $SU(2)$ R-symmetry rotation with the same coefficient as we found on $\psi$ in the tensor multiplet. 

The constraint equations (\ref{Co1}) are algebraic and can be solved at each point in space by the same argument as we used for the Poincare supersymmetry parameters. In flat space (\ref{Co1}) decomposes into the following constraints,
\bea
\bar\epsilon \h\Gamma^i \nu^{\alpha} &=& 0\cr
\bar\eta \h\Gamma^i \Gamma_M \nu &=& \bar\epsilon \h\Gamma^i \Gamma_M \rho\cr
\bar\eta \h\Gamma^i \rho &=& 0
\eea

Next we consider closure on the auxiliary fields,
\ben
\delta^2 F^{\alpha} &=& - i \bar\nu_f^{\alpha} \Gamma^M \Gamma^N \h\Gamma^i \Gamma_M \eta D_N\phi^i\label{ett}\\
&& - i \bar\nu_f^{\alpha} \h\Gamma^i\epsilon_f D^2_M\phi^i\label{tva}\\
&& - i \bar\nu_f^{\alpha} \Gamma^M D_M\nu_f^{\beta} F^{\beta}\label{tre}\\
&& - i \bar\nu_f^{\alpha} \Gamma^M \nu_f^{\beta} D_M F^{\beta}\label{fyra}\\
&& + 4 i \bar\nu_f^{\alpha} \Gamma^M \h\Gamma^i D_M \eta \phi^i\label{fem}\\
&& + 4 i \bar\nu_f^{\alpha} \Gamma^M \h\Gamma^i\eta D_M\phi^i\label{sex}
\een
We now analyze each line in turn. By using the identity $\Gamma^M \Gamma^N \Gamma_M = - 4 \Gamma^N$ we see that (\ref{ett}) cancels against (\ref{sex}). Line (\ref{tva}) vanishes by using the constraint (\ref{B1}). We decompose (\ref{tre}) in a symmetric and an antisymmetric piece
\bea
i \bar\nu_f^{\alpha} \Gamma^M D_M\nu_f^{\beta} F^{\beta} &=& i \bar\nu_f^{(\alpha} \Gamma^M D_M\nu_f^{\beta} F^{\beta)} + i \bar\nu_f^{[\alpha} \Gamma^M D_M\nu_f^{\beta]} F^{\beta}
\eea
and write the symmetric piece as
\bea
i \bar\nu_f^{(\alpha} \Gamma^M D_M\nu_f^{\beta)} F^{\beta} &=& - \frac{i}{2} (D^M S_M) F^{\alpha}
\eea
The antisymmetric piece gives an internal rotation,
\bea
i \bar\nu_f^{[\alpha} \Gamma^M D_M\nu_f^{\beta]} F^{\beta} &=& R^{\alpha\beta} F^{\beta}
\eea
in the 4d space of auxiliary fields. While this is a new sort of transformation, it is a symmetry of the theory. In the Lagrangian we shall have the term $F^{\alpha} F^{\alpha}$, which is invariant. Eq (\ref{fyra}) is the translation term,
\bea
- i S^M D_M F^{\alpha}
\eea
Eq (\ref{fem}) is vanishing by using (\ref{Co1}). We summarize the result as
\bea
\delta^2 F^{\alpha} &=& - i S^M D_M F^{\alpha} - 6 i \bar\epsilon_f \eta F^{\alpha} + R^{\alpha\beta} F^{\beta}
\eea
We can now recognize this as the closure relation that gives us the scaling dimension for $F^{\alpha}$ as $\triangle = 3$. 

We finally consider closure on the fermions
\bea
\delta^2 \chi &=& i \(\Gamma^M\h\Gamma^i \epsilon_f\bar\epsilon_f \h\Gamma^i - i\nu_f^{\alpha}\bar\nu_f^{\alpha}\Gamma^M\) D_M\chi\cr
&& - i \(\Gamma^M \h\Gamma^i \epsilon_f\bar\eta \Gamma_M \h\Gamma^i + 4 \h\Gamma^i \eta\bar\epsilon_f \h\Gamma^i\) \chi
\eea
The first line has exactly the same structure as we had in flat Poincare superspace, we just need to attach subscripts $f$ to the supersymmetry parameters and then we can borrow the flat space result. Let us therefore consider the second line. We use
\bea
\epsilon_f \bar\eta &=& c + c^{RS} \Gamma_{RS} + ...\cr
\eta \bar\epsilon_f &=& - c+ c^{RS} \Gamma_{RS} + ...
\eea
and we do not need to consider the dots, because of the identity $\h\Gamma^i \h\Gamma^{kl} \h\Gamma^i$. We then get that the second line is equal to 
\bea
- 5 i \bar\epsilon_f \eta \chi + \frac{i}{2} \bar\epsilon_f \Gamma_{RS} \eta \Gamma^{RS} \chi
\eea
which gives the scaling dimension $5/2$. We have no contribution from the other terms thanks to the identity $\h\Gamma^i \h\Gamma^{kl} \h\Gamma^i = 0$. Let us then summarize this,
\bea
\delta^2 \chi &=&  - i \L_S \chi - 5 i \bar\epsilon_f\eta \chi 
\eea
where things nicely combine into the Lie derivative
\bea
\L_S \chi &=& S^M D_M \chi + \frac{1}{4} D_M S_N \Gamma^{MN} \chi\cr
&=& S^M D_M \chi - \frac{1}{2} \bar\epsilon_f\Gamma_{MN}\eta \Gamma^{MN} \chi
\eea
This completes the closure compuation for Abelian gauge group.

\subsection{Non-Abelian gauge group and Poincare supersymmetry}
For the non-Abelian case, we have two options as far as Poincare supersymmetry concerns. Either we can couple the hypermultiplet to a $(1,0)$ vector multiplet, or we can couple it to a $(1,0)$ tensor multiplet. Here we will consider the latter case, and make the ansatz
\bea
\delta \phi^i &=& i \bar\epsilon\h\Gamma^i\chi\cr
\delta \chi &=& \Gamma^M \h\Gamma^i \epsilon D_M \phi^i + \Gamma^M \h\Gamma^i \epsilon [v_M,\sigma,\phi^i] + \nu^{\alpha} F^{\alpha}\cr
\delta F^{\alpha} &=& - i \bar\nu^{\alpha} \(\Gamma^M D_M \chi + \Gamma^M \h\Gamma^i [\psi,v_M,\phi^i] + \Gamma^M [\chi,v_M,\sigma]\)
\eea
where $\psi$ and $\sigma$ are the tensor multiplet fields whose supersymmetry variations we will take to be the on-shell variations. 

We may assume that we have the constraints (\ref{B1}), (\ref{B2}) and (\ref{B3}) satisfied. There is no way that these constraints can get modified by making the gauge group non-Abelian since the supersymmetry parameters do not carrying any gauge group index. 

We begin by closure on the scalars. We get
\bea
\delta^2 \phi^i &=& - i S^M D_M \phi^i - (\Lambda,\phi^i)
\eea
where the gauge parameter is
\bea
\Lambda &=& i S_M \<v^M,\sigma\>
\eea

Next we consider closure on the fermions, 
\ben
\delta^2 \chi &=& i \(\Gamma^M \h\Gamma^i \epsilon\bar\epsilon \h\Gamma^i + \nu^{\alpha}\bar\nu^{\alpha} \Gamma^M\) D_M \chi\label{C1}\\
&& + i \(\Gamma_N \h\Gamma^i \epsilon\bar\epsilon \Gamma^{NM} + \Gamma^M\h\Gamma^i \epsilon\bar\epsilon - \nu^{\alpha}\bar\nu^{\alpha} \Gamma^M \h\Gamma^i \) [\psi,v_M,\phi^i]\label{C2}\\
&& + i \(\Gamma^M\h\Gamma^i \epsilon\bar\epsilon \h\Gamma^i - \nu^{\alpha}\bar\nu^{\alpha}\) [\chi,v_M,\sigma]\label{C3}
\een
Line (\ref{C3}) gives the gauge transformation term
\bea
(\ref{C3}) &=& -(\Lambda,\chi)
\eea
with 
\bea
\Lambda &=& i S^M \<v_M,\sigma\>
\eea
We find that 
\bea
(\ref{C2}) &=& 0
\eea
by using a Fierz identity and the identities
\bea
-\Gamma_N \Gamma_Q \Gamma^{NM} - \Gamma^M\Gamma_Q - 4\Gamma_Q\Gamma^M &=& 0\cr
-\Gamma_N \Gamma_{RST} \Gamma^{NM} - \Gamma^M \Gamma_{RST} &=& 0
\eea
Line (\ref{C1}) is of the same form as for the Abelian case, and thus we have got the off-shell closure relation,
\bea
\delta^2 \chi &=& - i S^M D_M \chi - (\Lambda,\chi)
\eea

Finally we turn to closure on $F^{\alpha}$. By using the on-shell variations for the tensor multiplet fields $\sigma$ and $\psi$, we get
\ben
\delta F^{\alpha} &=& - \frac{i}{2} \bar\nu^{\alpha} \Gamma^{MN} \h\Gamma^i \epsilon (F_{MN},\phi^i)\label{F1}\\
&& - i\bar\nu^{\alpha} \h\Gamma^i \Gamma^{MN} \epsilon [D_M v_N,\sigma,\phi^i]\label{FSurvives}\\
&& - i \bar\nu^{\alpha} \Gamma^{MN} \h\Gamma^i \epsilon [v_N,D_M\sigma,\phi^i]\label{F2}\\
&& - i \bar\nu^{\alpha} \Gamma^{MN} \h\Gamma^i\epsilon [v_N,\sigma,D_M\phi^i]\label{F3}\\
&& - i \bar\nu^{\alpha} \Gamma^M \nu^{\beta} D_M F^{\beta}\label{F4}\\
&& + \frac{i}{12} \bar\nu^{\alpha} \Gamma_M \Gamma^{PQR} \h\Gamma^i \epsilon [H_{PQR},v^M,\phi^i]\label{F5}\\
&& - i \bar\nu^{\alpha} \Gamma^{MN} \h\Gamma^i \epsilon [v_M,D_N\sigma,\phi^i]\label{F6}\\
&& - i \bar\nu^{\alpha} \Gamma^M \nu^{\beta} [v_M,\sigma,F^{\beta}]\label{F7}\\
&& + i \bar\nu^{\alpha} \Gamma^{MN} \h\Gamma^i \epsilon [v_M,D_N\phi^i,\sigma]\label{F8}\\
&& + \frac{i}{2} \bar\nu^{\alpha} \Gamma_M \h\Gamma^i \Gamma^Q \h\Gamma^{kl} \epsilon [v^M,\phi^i,[v_Q,\phi^k,\phi^{\l}]]\label{F9}\\
&& - i \bar\nu^{\alpha} \Gamma^M \Gamma^N \h\Gamma^i \epsilon [v_M,\sigma,[v_N,\sigma,\phi^i]]\label{F10}
\een
Lines (\ref{F4}) and (\ref{F7}) give the translation and the gauge transformation respectively,
\bea
\delta^2 F^{\alpha} &=& - i S^M D_M F^{\alpha} - (\Lambda,F^{\alpha})
\eea
which is the off-shell closure relation that we want. It now remains to show that all the remaining terms vanish. Lines (\ref{F1}) and (\ref{F5}) cancel if we assume that
\bea
F_{MN} &=& \<H_{MNP},v^P\>
\eea
Lines (\ref{F2}) and (\ref{F8}) cancel. Lines (\ref{F3}) and (\ref{F8}) cancel. Each of the double commutators in lines (\ref{F9}) and (\ref{F10}) vanish by themselves, by applying the fundamental identity in the form
\bea
[\alpha,\beta,[\gamma,\delta,\bullet]] - [\gamma,\delta,[\alpha,\beta,\bullet]] &=& [[\alpha,\beta,\gamma],\delta,\bullet] + [\gamma,[\alpha,\beta,\delta],\bullet]
\eea
and by assuming that 
\bea
\<v_M,v_N\> &=& 0
\eea
Line (\ref{FSurvives}) survives, unless we impose the constraint 
\bea
\Gamma^{MN}\epsilon D_M v_N &=& 0
\eea 
In the next subsection we will see that the left hand side is part of the Lie derivative $\L_v \epsilon$, which we can not see now as we assume that $\epsilon$ is a constant Poincare supersymmetry parameter.

This is not end of the story since we have further terms that are bilinear in fermionic fields. These terms are
\bea
(\delta^2 F^i)_{bilin} &=& i \bar\nu \h\Gamma^i \Gamma^M (\delta A_M,\chi) - i \bar\nu \h\Gamma^i \Gamma^M \h\Gamma^j [\psi,v_M,\delta \phi^j] - i \bar\nu\h\Gamma^i \Gamma^M [\chi,v_M,\delta\sigma]
\eea
These terms are algebraic in the sense that they do not involve any derivatives, so they share exactly the same gamma matrix and Fierz identity structures as the corresponding terms that arise in the 5d closure computation in \cite{Hosomichi:2012ek}, and so these terms shall cancel out here as well. This relation to \cite{Hosomichi:2012ek} becomes clear if one chooses the Abelian vector field $v_M$ to lie in say the 0-th direction and the generalization to arbitrary vector field then follows by the fact that this contribution is algebraic, together wih the fact that $\delta A_0 \sim \Gamma_{00} = 0$ so that we are effectively left with only the 5d contribution. In appendix $B$ we show how to translate our spinor language into the spinor language of \cite{Hosomichi:2012ek}.

By taking away the auxiliary field $F^{\alpha}$ and closing these variations on-shell, we obtain the fermionic equation of motion
\bea
\Gamma^M D_M \chi + \Gamma^M \h\Gamma^i [\psi,v_M,\phi^i] + \Gamma^M [\chi,v_M,\sigma] &=& 0
\eea
We see that it is this equation of motion that reappears in $\delta F^{\alpha}$ when we close supersymmetry off-shell.

\subsection{Non-Abelian gauge group and superconformal symmetry}
We make the ansatz
\bea
\delta \phi^i &=& i \bar\epsilon_f\h\Gamma^i\chi\cr
\delta \chi &=& \Gamma^M \h\Gamma^i \epsilon_f D_M \phi^i + \Gamma^M \h\Gamma^i \epsilon_f [v_M,\sigma,\phi^i] + \nu^{\alpha} F^{\alpha} - 4 \h\Gamma^i \eta \phi^i\cr
\delta F^{\alpha} &=& - i \bar\nu^{\alpha} \(\Gamma^M D_M \chi + \Gamma^M \h\Gamma^i [\psi,v_M,\phi^i] + \Gamma^M [\chi,v_M,\sigma]\)
\eea
New type of terms in the closure computation arise when 3-brackets mix and curvature corrections mix. Such a mixing does not arise when we vary $\phi^i$ or $\chi$ twice, but it arises at many places when we vary $F^{\alpha}$ twice. 

We can use the previous computations to argue that we have closure on $\phi^i$ with the gauge parameter $\Lambda = i S^M \<v_M,\sigma\>$. When we vary $\phi^i$ twice we get a 3-bracket term plus a curvature term. For both terms we have already shown closure, although for the 3-bracket term we now need to substitute $\epsilon$ in the previous computation with $\epsilon_f$ here. For this case, a direct computation is also very easy to carry out. In the same way we can argue that we have closure on $\chi$ up to the same gauge transformation. Although here we need to vary some fields that sit inside 3-brackets, all those fields are bosonic and for the variations of the bosonic fields there are no curvature corrections and so we find no mixing between curvature corrections and 3-brackets, and so we have closure on $\chi$. 

New contributions arise when we make a second supersymmetry variation of $\delta F^{\alpha}$. We find two types of new terms. One type of new terms arise from the 3-brackets that contain a fermion field in one of the entries. The variation of fermion fields have curvature correction terms, which produce new terms. They other type of contribution arises from when the derivative hits $\epsilon_f$. All such terms where encountered in the case of Abelian gauge group, except for the 3-bracket in $\delta\chi$ that now produces the second type of new term. We thus only need to examine these new kind of terms. They are given by
\bea
(\delta^2 F^{\alpha})_{new} &=& - \(i \bar\nu^{\alpha} \Gamma^M D_M (\delta \chi)_{comm}\)_{new}\cr
&&  - i \bar\nu^{\alpha} \Gamma^M \h\Gamma^i [(\delta\psi)_{curv},v_M,\phi^i] - i \bar\nu^{\alpha} \Gamma^M [(\delta\chi)_{curv},v_M,\sigma]
\eea
where 
\bea
(\delta \chi)_{comm} &=&  \Gamma^M \h\Gamma^i \epsilon_f [v_M,\sigma,\phi^i]\cr
(\delta \chi)_{curv} &=& - 4 \h\Gamma^i \eta \phi^i\cr
(\delta \psi)_{curv} &=& 4 \eta \sigma
\eea
In the first term we will pick out the contributions when the derivative hits $\epsilon_f$ and $v_M$, since previously we have obtained the contributions from when the derivative hits the other fields. This is why we indicate this by the subscript $new$ in the first term. 

Plugging in these variations, we get
\bea
(\delta^2 F^{\alpha})_{new} &=& - 4 i \bar\nu^{\alpha} \h\Gamma^i (D_M \epsilon_f) [v^M,\sigma,\phi^i] - i\bar\nu^{\alpha} \h\Gamma^i \Gamma^{MN} \epsilon_f [D_M v_N,\sigma,\phi^i]
\eea
These two terms nicely combine into 
\bea
(\delta^2 F^{\alpha})_{new} &=& - 4 i \bar\nu^{\alpha} \h\Gamma^i [\L_v \epsilon_f,\sigma,\phi^i]
\eea
where
\bea
\L_v \epsilon_f &=& v^M D_M \epsilon_f + \frac{1}{4} D_M v_N \Gamma^{MN} \epsilon_f
\eea
To get closure on $F^{\alpha}$ we need to impose the constraint 
\ben
\L_v \epsilon_f &=& 0\label{Le}
\een
that we saw already for the tensor multiplet. So it is not a surprise to also find this constraint again here for the hypermultiplet. While Lie derivatives of bosonic fields can be defined for any vector field, the Lie derivative on a spinor can be introduced only if that vector field is a Killing vector field. So the constraint (\ref{Le}) implies in particular that the 6d manifold has at least one isometry direction and that $\L_v g_{MN} = 0$.

\section{Solving two constraints in Lorentzian 3-algebra}
We will now restrict ourselves to Lorentzian 3-algebras \cite{Awata:1999dz}, \cite{Gomis:2008uv}, \cite{Benvenuti:2008bt}, \cite{Ho:2008ei} for which we have generators $T^{\pm}$ together with Lie algebra generators $T^a$ that satisfy 
\bea
[T^a,T^b] &=& f^{ab}{}_c T^c
\eea
The generator $X = X_a T^a + X_+ T^+ + X_- T^-$ has the norm
\bea
X\cdot X &=& h^{ab} X_a X_b + 2 X_+ X_-
\eea
The 3-algebra relations are
\bea
[T^a,T^b,T^+] &=& f^{ab}{}_c T^c\cr
[T^a,T^b,T^-] &=& 0\cr
[T^a,T^b,T^c] &=& f^{abc}
\eea
where $f^{abc} = f^{ab}{}_d h^{dc}$ and $h^{ab}$ is a gauge invariant metric, so that $f^{abc}$ is totally antisymmetric. Similarly, let us also introduce the 3-algebra structure constants $f^{ab+}{}_c = f^{ab}{}_c$ which have the properties that $f^{abc+}$ is totally antisymmetric in all four entries $a,b,c,+$. We solve the Abelian constraint by taking 
\bea
v^M &=& V^M T^+
\eea
where we use the notation $V^M := v^M_+$. Since all interaction terms are mediated by the coupling vector $V^M$, we see that 3-algebra valued fields multiplying $T^{\pm}$ will amount to two free tensor multiplets. Let us from now on discard these, and only retain the interacting 3-algebra valued fields, which are those that take values in the Lie algebra generated by the $T^a$'s. Thus we expand any non-Abelian 3-algebra valued field as $\varphi = \sigma_a T^a$ where we discard the free field part $\varphi_+ T^+ + \varphi_- T^-$. Let us now clarify the map from the 3-algebra notation to the Lie algebra notiation. First we notice that if we have two fields $\varphi_1$ and $\varphi_2$ whose free field parts have been discarded, the 3-algebra inner product reduces to the inner product on the Lie algebra,
\bea
\varphi_1 \cdot \varphi_2 &=& h^{ab}\varphi_{1a} \varphi_{2b}
\eea
and so we do not need to use a separate notation for the inner product on the Lie algebra. Let us next consider the covariant derivative. We expand out $(A_M,\sigma) = [B_{MN},v^N,\sigma] = B_{MN,a}v^N_+ [T^a,T^+,\sigma] = - B_{MN,a} V^N [T^a,\sigma]$. Thus if we define 
\bea
A_M = A_{M,a} T^a = B_{MN,a} V^N T^a
\eea
then we will have the covariant derivative
\bea
D_M \sigma = \partial_M \sigma - [A_M,\sigma]
\eea
and we will have the constraint
\bea
F_{MN,a} &=& H_{MNP,a} V^P 
\eea
To solve this constraint, we make the ansatz
\bea
H_{MNP,a} &=& F_{MN,a}\kappa_{P} + F_{NP,a}\kappa_M + F_{PM,a}\kappa_N + C_{MNP,a}
\eea
where $C_{MNP,a} V^P = 0$. If we plug this ansatz into the constraint, it reduces to
\bea
F_{MN,a} &=& F_{MN,a}\kappa_P V^P + 2 F_{P[M,a}\kappa_{N]} V^P
\eea
This leads to the constraints
\ben
\kappa_{[M} V_{N]} &=& 0\label{L1}\\
\kappa_P V^P &=& 1\label{L2}\\
F_{MN,a} V^N &=& 0\label{L3}
\een
The constraints (\ref{L1}) and (\ref{L2}) are solved by taking 
\bea
\kappa_M &=& \frac{V_M}{g^2},\cr
g^2 &=& V^M V_M
\eea
The constraint (\ref{L3}) says that
\bea
\L_V A_M + [\Lambda,A_M] &=& 0
\eea
where $\L_V A_M = V^N \partial_N A_M + (\partial_M V^N) A_N$ and $\Lambda = V^M A_M$. We will impose the gauge fixing condition that puts $\Lambda = 0$. 

The on-shell supersymmetry variations for the $(1,0)$ tensor multiplet in the Lie algebra formulation read 
\ben
\delta A_M &=& i \bar\epsilon_f \Gamma_{MN} \psi V^N\cr
\delta H_{MNP} &=& 3 i \bar\epsilon_f \Gamma_{[NP} D_{M]}\psi - 3 i \bar\eta\Gamma_{MNP} \psi + i \bar\epsilon_f\Gamma_{MNPQ} [\psi,\sigma]V^Q\cr
\delta \sigma &=& - i \bar\epsilon_f \psi\cr
\delta \psi &=& \frac{1}{12} \Gamma^{MNP} \epsilon_f H_{MNP} + \Gamma^M \epsilon_f D_M \sigma + 4 \eta \sigma \label{Lambert}
\een
We shall remember the sign flip in the covariant derivatives, $D_M \varphi = \partial_M \varphi - [A_M,\varphi]$ in the Lie algebra formulation. 

\section{Tensor multiplet without the tensor field}
When we close supersymmetry variations on $H_{MNP}$, we find a modified Bianchi identity with the following structure \cite{Lambert:2010wm}\footnote{Here $\lambda$ is a constant that can be determined with some efforts, as was done in \cite{Lambert:2010wm}.}
\ben
D_{[Q} H_{MNP]} &=& \frac{1}{4} \epsilon_{MNPQ}{}^{RS} [\sigma,D_R \sigma] V_S + \frac{i\lambda}{8} \epsilon_{MNPQ}{}^{RS} [\bar\psi,\Gamma_R\psi]V_S\label{mBianchi}
\een
If we contract by $V^P$, the left-hand side of this modified Bianchi identity trivializes,
\bea
\(4 D_{[Q} H_{MNP]}\) V^P = 3 D_{[Q} F_{MN]} - \L_V H_{QMN} = 0
\eea
and so does the right-hand side as well, thanks to the Abelian constraint, or more simply, $V_{[M} V_{N]} = 0$.

Since only the selfdual part of $H_{MNP}$ enters in the supersymmetry variation of $\psi$, one may expect one can use only the components $F_{MN,a} = H_{MNP,a} V^P$ since the other components are related by selfduality. Indeed we find that the following non-Abelian $(2,0)$ tensor multiplet Lagrangian 
\bea
\L &=& - \frac{1}{4g^2} F^{MN} \cdot F_{MN} - \frac{1}{2} D^M \phi^A \cdot D_M \phi^A + \frac{i}{2} \bar\Psi \cdot \Gamma^M D_M \Psi - \frac{R}{10} \phi^A \cdot \phi^A\cr
&& - \frac{i}{2} \bar\Psi \cdot \Gamma^M \h\Gamma^A [\Psi,\phi^A] V_M - \frac{g^2}{4} [\phi^A,\phi^B]\cdot [\phi^A,\phi^B]
\eea
has the variation 
\bea
\delta \L &=& - \frac{i}{2} \bar\Psi \Gamma^{MNPQ} \epsilon \cdot D_M \(F_{NP} \kappa_Q\)\cr
&& + \frac{i}{2g^2} \L_V \(\bar\Psi \Gamma^{MN} \epsilon\)\cdot F_{MN}\cr
&& - \(\frac{i}{2} \L_V \(\bar\Psi \h\Gamma^{AB} \epsilon\) + 2 i \bar\Psi \h\Gamma^{AB} \L_V \epsilon\)\cdot [\phi^A,\phi^B]
\eea
under the $(2,0)$ supersymmetry variations
\bea
\delta \phi^A &=& i \bar\epsilon_f \h\Gamma^A \Psi\cr
\delta \Psi &=& \Gamma^M \h\Gamma^A \epsilon_f D_M \phi^A + \frac{1}{2} \Gamma^{MNP} \epsilon_f F_{MN} \kappa_P - 4 \h\Gamma^A \eta \phi^A - \frac{1}{2} \Gamma^M \h\Gamma^{AB} \epsilon_f [\phi^A,\phi^B] V_M\cr
\delta A_M &=& i \bar\epsilon_f \Gamma_{MN} \Psi V^N
\eea
where 
\bea
g^2 &:=& V^M V_M
\eea
Thus the action becomes supersymmetric if we impose the constraints
\bea
\L_V ({\mbox{fields}}) &=& 0\cr
\L_V \epsilon &=& 0
\eea
and if we add a graviphoton term $\L_{graviphoton}$ whose variation is
\bea
\delta \L_{graviphoton} &=& \frac{i}{2} \bar\psi \Gamma^{MNPQ} \epsilon \cdot D_M \(F_{NP} \kappa_Q\)
\eea
In this graviphoton term where all vector indices are antisymmetrized, no Christoffel symbol survives and in fact the term is topological. However, this graviphoton term can be explicitly expressed only if we switch to the 5d langauge. To this end, we start by writing the 6d metric in the form
\ben
ds^2 &=& g(x^{\mu})^2 \(dy + \kappa_{\mu} dx^{\mu}\)^2  + G_{\mu\nu}(x^{\mu}) dx^{\mu} dx^{\nu}\label{circlebundle}
\een
We identify our Killing vector field as
\bea
V &=& \frac{\partial}{\partial y}
\eea
and $\kappa_M = (\kappa_{\mu},1)$ where $x^M = (x^{\mu},y)$. Then we have
\bea
\delta \L_{graviphoton} &=& \frac{i}{2} \bar\psi \Gamma^{\mu\nu\lambda\tau} \epsilon \cdot D_{\mu} \(F_{\nu\lambda} \kappa_{\tau}\)\cr
&& +  \frac{i}{2} \bar\psi \Gamma^{\mu\nu\lambda y} \epsilon \cdot D_{\mu} F_{\nu\lambda}\cr
&& +  \frac{i}{2} \bar\psi \Gamma^{y\nu\lambda\tau} \epsilon \cdot D_y \(F_{\nu\lambda} \kappa_{\tau}\)
\eea
The second term vanishes by the Bianchi identity $D_{[\mu} F_{\nu\lambda]} = 0$. There are no terms involving $F_{M y} = F_{MN} V^N$ since this is zero by the constraints $\L_V($fields$) = 0$ together with the gauge fixing condition $A_y = A_M V^M = 0$. Finally we notice that $\L_V (F_{\nu\lambda} \kappa_{\tau}) = \partial_y (F_{\nu\lambda} \kappa_{\tau})$ but we got $D_y$ in place of $\partial_y$ in the variation above. But then we notice that the metric plays no role in this variation which is a topological term, so we may right away take away the Christoffel symbols from all the covariant derivatives. Second, $A_y = 0$. After this, we have $D_y = \partial_y$ and thus
\bea
\delta \L_{graviphoton} &=& \frac{i}{2} \bar\psi \Gamma^{\mu\nu\lambda\tau} \epsilon \cdot D_{\mu} \(F_{\nu\lambda} \kappa_{\tau}\)
\eea
This is the variation of the graviphoton term \cite{Linander:2011jy}
\bea
\L_{graviphoton} &=& \frac{1}{8} \epsilon^{\mu\nu\lambda\kappa\tau} F_{\mu\nu} \cdot F_{\lambda\kappa} \kappa_{\tau}
\eea

Since in the end we want to obtain off-shell supersymmetry, we will split this Lagrangian into a sum
\bea
\L &=& \L_T + \L_H
\eea
consisting of one tensor multiplet part
\bea
\L_T &=& - \frac{1}{4g^2} F^{MN} \cdot F_{MN} - \frac{1}{2} D^M \sigma\cdot D_M \sigma + \frac{i}{2} \bar\psi \cdot \Gamma^M D_M \psi - \frac{R}{10}\sigma\cdot \sigma\cr
&& + \frac{i}{2} \bar\psi \cdot \Gamma^M [\psi,\sigma] V_M 
\eea
and one hypermultiplet part
\bea
\L_H &=& - \frac{1}{2} D^M \phi^i \cdot D_M \phi^i + \frac{i}{2} \bar\chi \cdot \Gamma^M D_M \chi - \frac{R}{10} \phi^i \cdot \phi^i\cr
&& - \frac{i}{2} \bar\chi \cdot \Gamma^M [\chi,\sigma] V_M - i \bar\psi \cdot \Gamma^M \h\Gamma^i [\chi,\phi^i] V_M \cr
&& - \frac{g^2}{2} [\phi^i,\sigma]\cdot [\phi^i,\sigma] - \frac{g^2}{4} [\phi^i,\phi^j]\cdot [\phi^i,\phi^j]
\eea
Corresponding supersymmetry variations are 
\bea
\delta \sigma &=& - i \bar\epsilon_f \psi\cr
\delta A_M &=& i \bar\epsilon_f \Gamma_{MN} \psi V^N\cr
\delta \psi &=& \frac{1}{2} \Gamma^{MNP} \epsilon_f F_{MN}\kappa_P + \Gamma^M \epsilon_f D_M \sigma - \frac{1}{2} \Gamma^M \h\Gamma^{ij} \epsilon_f [\phi^i,\phi^j]V_M + 4 \eta \sigma
\eea
and 
\bea
\delta \phi^i &=& i\bar\epsilon_f\h\Gamma^i\chi\cr
\delta \chi &=& \Gamma^M \h\Gamma^i \epsilon_f D_M \phi^i - \Gamma^M \h\Gamma^i \epsilon_f [\phi^i,\sigma]V_M - 4 \h\Gamma^i \eta \phi^i
\eea
As we now have reduced supersymmetry from $(2,0)$ down to $(1,0)$, we should ask ourselves if there could be a free parameter which is not completely fixed by $(1,0)$ supersymmetry. By making the most general ansatz with arbitrary coefficients, we have shown that $(1,0)$ supersymmetry fixes all the coefficients uniquely. Hence one $(1,0)$ tensor multiplet coupled to one adjoint $(1,0)$ hypermultiplet will automatically become $(2,0)$ supersymmetric. There is no free parameter, other than the vector field $V_M$ that has to be a Killing vector field, but which is otherwise our free choice. To get a genuinely $(1,0)$ supersymmetric theory, we may consider just the $(1,0)$ tensor multiplet in isolation, with no hypermultiplet, or we may consider several hypermultiplets in different representations. Our result is in agreement with that of \cite{Chen:2013wya}, although in that reference one free parameter $b$ appeared. However, that parameter can be absorbed into the hypermultiplet fields by a simple field rescaling, which means that this is not a genuine parameter that parametrizes physically different theories. We may also turn on supergravity background fields, such as the R-gauge field, which can have the effect of both reducing the supersymmetry down to $(1,0)$ as well as bringing in an extra free parameter. 

Before addressing the invariance of the action, we consider the closure relations for the $(1,0)$ tensor multiplet. Let us drop the coupling term $ - \frac{1}{2} \Gamma^M \h\Gamma^{ij} \epsilon_f [\phi^i,\phi^j]V_M$ to the hypermultiplet for simplicity. (Keeping it, will only have the effect of modifying the tensor multiplet fermionic equation of motion by an extra coupling term to the hypermultiplet.) Then for the scalar we get the same kind of closure relation as before,
\bea
\delta^2 \sigma &=& -i S^M D_M \sigma - 4 i \bar\epsilon_f \eta \sigma
\eea
For the gauge field $A_M$, we get
\bea
\delta^2 A_M &=& - i S^T F_{TM} + D_M \Lambda + i \L_V\(S_M \sigma\)\cr
\Lambda &=& - i S_N V^N \sigma
\eea
by using 
\bea
F_{MN} V^N &=& 0\cr
\kappa_{[M} V_{N]} &=& 0
\eea
For the fermions we get\footnote{The dots contain the R-rotation, the scaling dimension and all those terms we saw before in the 3-algebra language.}
\bea
\delta^2 \psi &=& - i \L_S \psi - [\psi,\Lambda] + ...\cr
&& + \frac{3i}{8} S^Q \(\Gamma^Q - 2 V_Q \kappa_P \Gamma^P\) \(\Gamma^M D_M \psi + \Gamma^M[\psi,\sigma]V_M\)
\eea
with 
\bea
\Lambda &=& - i S^M V_M\sigma
\eea
To see this requires a small trick where we remove a certain Lie derivative $\L_V$ by using constraints. The tricky term is
\bea
(\delta^2 \psi)_{tricky} &=& \frac{1}{2} \Gamma^{MNP} \epsilon_f \delta F_{MN} \kappa_P
\eea
We use the identity
\bea
\delta F_{MN} &=& \delta H_{MNP} V^P - i \L_V\(\bar\epsilon_f \Gamma_{MN} \psi\)
\eea
and the trick is to note that by our constraints, we can drop the second term. Then we get
\bea
(\delta^2 \psi)_{tricky} &=& \frac{1}{2} \Gamma^{MNP} \epsilon_f \delta H_{MNQ} V^Q \kappa_P
\eea
Expanding it out, we get a sum of two terms
\bea
(\delta^2 \psi)_{tricky} &=& (\delta^2\psi)_{tricky,1} + (\delta^2 \psi)_{tricky,2} 
\eea
where
\bea
(\delta^2\psi)_{tricky,1} &=& \frac{3i}{2} \Gamma^{MNP} \epsilon_f \bar\epsilon_f\Gamma_{[NQ} D_{M]} \psi V^Q \kappa_P\cr
(\delta^2\psi)_{tricky,2} &=& -\frac{3i}{2} \Gamma^{MNP} \epsilon_f \bar\eta \Gamma_{MNQ} \psi V^Q \kappa_P
\eea
It is the term $(\delta^2\psi)_{tricky,2}$, that along with some other terms, will give us the correct scaling dimension $\triangle = 5/2$ for the fermion. For this part, the computation is completely analogous to our previous computation in 3-algebra language. 

Let us now study the first term. This term can be separated into two terms
\bea
(\delta^2 \psi)_{tricky,1} &=& (\delta^2 \psi)_{tricky,1a} + (\delta^2 \psi)_{tricky,1b}
\eea
where 
\bea
(\delta^2 \psi)_{tricky,1a} &=& i \Gamma^{MNP} \epsilon_f \bar\epsilon_f \Gamma_{NQ} D_M \psi V^Q \kappa_P\cr
(\delta^2 \psi)_{tricky,1b} &=& \frac{i}{2} \Gamma^{MNP} \epsilon_f \bar\epsilon_f \Gamma_{MN} D_Q \psi V^Q \kappa_P
\eea
The first term $(\delta^2 \psi)_{tricky,1a}$ is what we would like to refer to as a basic term. That is, a term that also would arise in 5d Poincare supersymmetry variations. It will combine with other equally basic terms into a basic variation,
\bea
(\delta^2 \psi)_{basic} &=& i \(\Gamma^{MNP} \epsilon_f \bar\epsilon_f \Gamma_{NR} V^R \kappa_P - \Gamma^M \epsilon_f \bar\epsilon_f\) D_M \psi - i \Gamma^M \epsilon_f \bar\epsilon_f \Gamma_{MN} [\psi,\sigma] V^N
\eea
Let us now compute the contribtion from the first term in the Fierz identity,
\bea
\epsilon_f \bar\epsilon_f &=& c_Q \Gamma^Q + ...
\eea
We note the following identity
\bea
\Gamma^{MNP} \Gamma^Q \Gamma_{NR} V^R \kappa_P &=& 6 V^Q \kappa^M - 6 \Gamma^P \Gamma^M \kappa_P V^Q + 6 \kappa^Q V^M - 2 \Gamma^Q \Gamma^P \kappa_P V^M \cr
&&- 6 g^{MQ} + 2 \Gamma^Q \Gamma^M
\eea
Then we get
\ben
&&- i S^Q D_Q \psi \cr
&&+ \frac{3i}{8} S^Q\(\Gamma_Q - 2 V^Q \kappa_P \Gamma^P\) \(\Gamma^M D_M \psi + \Gamma^M[\psi,\sigma]V_M\)\cr
&&+ \frac{i}{8} S_Q\(12 \kappa^Q - 2 \Gamma^Q \Gamma^P \kappa_P\) V^M D_M \psi\label{basic}
\een
On the other hand, we get
\bea
(\delta^2 \psi)_{tricky,1b} &=& \frac{ic_Q}{2} \Gamma^{MNP} \Gamma^Q \Gamma_{MN} D_R \psi V^R \kappa_P\cr
&=& - \frac{i}{8} S_Q \(12 \kappa^Q - 2 \Gamma^Q \Gamma^P \kappa_P\) V^M D_M \psi
\eea
precisely canceling the second line in (\ref{basic}).

Let us now turn to the invariance of the tensor multiplet Lagrangian. The most nontrivial part is the one that involves $F_{MN}$. Those variations come from two places,
\bea
(\delta \L_T)_1 &:=& \delta \(- \frac{1}{4g^2} F^{MN} \cdot F_{MN}\) \cr
&=& i\bar\epsilon_f \Gamma_{NP} \psi V^P D_M \(\frac{1}{g^2} F^{MN}\)
\eea
and\footnote{Here we do not include the term that we get from letting $D_M$ act on $\epsilon_f$.} 
\bea
(\delta \L_T)_2 := - i \bar\psi \Gamma^M D_M \(\delta \psi\)_{F_{MN}} &=& - \frac{i}{2} \bar\psi \Gamma^M \Gamma^{RST} \epsilon_f D_M \(F_{RS} \kappa_T\)
\eea
Now we use $\kappa_M = V_M/g^2$ and 
\bea
\Gamma^M \Gamma^{RST} &=& 3 g^{M[R} \Gamma^{ST]} + \Gamma^{MRST} 
\eea
and we get
\bea
(\delta \L_T)_2 &=& - i \bar\epsilon_f \Gamma_{NP} \psi D_M \(\frac{1}{g^2} F^{MN}\) V^P\cr
&& + \frac{i}{2} \L_V\(\bar\epsilon_f \Gamma_{NP} \psi\) \frac{1}{g^2} F^{NP}\cr
&& - \frac{i}{2} \bar\psi \Gamma^{MRST} \epsilon_f D_M \(F_{RS} \kappa_T\)
\eea
We put the Lie derivative to zero by constraints on $\epsilon_f$ and $\psi$. We cancel the first term against $(\delta \L_T)_1$ and we are left with the last term
\bea
(\delta \L_T)_1 + (\delta \L_T)_2 &=& -\frac{i}{2} \bar\psi \Gamma^{MRST} \epsilon_f D_M\(F_{RS} \kappa_T\)
\eea
This variation is canceled by adding $\L_{graviphoton}$ to the Lagrangian, as we have discussed above. 

Let us next illustrate the absence of a modified Bianchi identity. This can be seen from varying the following terms in the $(1,0)$ tensor multiplet Lagrangian
\bea
\L_T &=& ... - \frac{1}{2} D^M \sigma \cdot D_M \sigma + \frac{i}{2} \bar\psi \cdot \Gamma^M D_M \psi + \frac{i}{2} \bar\psi \cdot \Gamma^M [\psi,\sigma] V_M
\eea
Retaining only the terms that could be of relevance to get a possible modified Bianchi identity, we find 
\bea
\delta \L_{T} &=& ... - \frac{i}{2} \bar\psi \Gamma^{MRST} \epsilon D_M \(F_{RS}\kappa_T\)\cr
&& + (1-1) i \bar\epsilon\Gamma^{MN} \psi  \cdot [D_M \sigma,\sigma]V_N\cr
&& - \frac{1}{2} \bar\epsilon \Gamma_N \Gamma_M \psi \cdot [\bar\psi,\Gamma^M\psi] V^N + \frac{1}{2}(1-1)\bar\epsilon\psi\cdot [\bar\psi,\Gamma_N\psi]V^N 
\eea
Here all brackets act on Lie-algebra indices only, so for example $[\psi,\Gamma^M \psi] := \bar\psi^a \Gamma^M \psi^b [T_a,T_b]$. By using the Fierz identity
\bea
\psi^a \bar\psi^b &=& - \frac{1}{8} \bar\psi^b \Gamma^M \psi^a P_+ \h P_-\Gamma_M - \frac{1}{96} \bar\psi^b \Gamma^{MNP} \h\Gamma^{ij} \psi^a \Gamma_{MNP} \h\Gamma_{ij}
\eea
we get the identity 
\bea
\Gamma_N \Gamma_M \psi^a \bar\psi^b \Gamma^M \psi^c f_{abc} &=& -\frac{1}{2} \Gamma_N \Gamma_M \psi^a \bar\psi^b \Gamma^M \psi^c f_{abc}
\eea
which means that this cubic term is zero. This computation is analogous with that of SYM theory \cite{Brink:1976bc}. Thus we see that all the terms that could have potentially have contributed to a modified Bianchi identity, all either cancel out (as we have indicated by writing $1-1$), or vanishes identically, and the remaining term is canceled by the variation of $\L_{graviphoton}$. If the terms had added up (so that instead of $1-1$ we had had $1+1$), then we would have had a modified Bianchi identity. 

We are left with checking many more terms in the Lagrangian, but most of those terms are analogous with that of 5d SYM theory. The new ingredient here is superconformal symmetry. In the appendix we present a detailed derivation for superconformal symmetry of the actions when the gauge group is Abelian. The remaining contributions that are nonstandard are coming from variations that mix commutators and the superconformal added contribtion $\delta' \psi = 4 \eta \sigma$. There is only one such a nonstandard contribution. This is coming from the variation of the commutator,
\bea
- i \bar\psi \cdot \Gamma^M [\delta'\psi,\sigma] V_M
\eea
but this vanishes because of $[\sigma,\sigma] = 0$.

\subsection{Taking the tensor multiplet off-shell}
We have not been able to take the Abelian tensor multiplet off-shell. However, the non-Abelian\footnote{By this we mean no $U(1)$ factors in the gauge group.} tensor multiplet should be, as we have argued, an uplift from the 5d vector multiplet to 6d, and it is a standard result that the 5d vector multiplet can be taken off-shell. So we should expect that its 6d uplift where we impose the constraints $\L_V = 0$ on everything, also has an off-shell generalization. 

These off-shell supersymmetry variations should be given by\footnote{For maximal simplicity, we have dropped the coupling to hypermultiplet fields again.}
\bea
\delta \sigma &=& - i \bar\epsilon_f\psi\cr
\delta A_M &=& i \bar\epsilon_f \Gamma_{MN} \psi V^N\cr
\delta \psi &=& \frac{1}{2} \Gamma^{MNP} \epsilon_f F_{MN} \kappa_P + \Gamma^M \epsilon_f D_M \sigma + 4 \eta \sigma + \frac{1}{2} \h\Gamma^{ij} \Gamma^M \epsilon_f D^{ij} \kappa_M\cr
\delta D^{ij} &=& - \frac{i}{2} \bar\epsilon_f \Gamma^P \h\Gamma^{ij} \(\Gamma^M D_M \psi + \Gamma^M [\psi,\sigma] V_M\)  V_P
\eea
where $D^{ij}$ are three selfdual auxiliary fields with scaling dimensions $\Delta = 2$. That scaling dimension can be read off from $\delta^2 D^{ij} = ... - 4i \bar\epsilon_f \eta D^{ij}$, but we can also see that the scaling dimensions
\bea
[\epsilon_f] &=& L^{1/2}\cr
[V_M] &=& L\cr
[\kappa_M] &=& L^{-1}\cr
[\psi] &=& L^{-5/2}\cr
[D^{ij}] &=& L^{-2}\cr
[\sigma] &=& L^{-2}\cr
[D_M] &=& L^{-1}
\eea
are consistent with these variations. These variations should close off-shell on the generators of the conformal group plus an R-symmetry rotation that acts on $D^{ij}$ and $\psi$. The non-Abelian generalization for the Poincare part of the supersymmetry should hold since we know that the existence of off-shell supersymmetry for the 5d vector multiplet and this is essentially just a trivial 6d uplift of that. It remains to check the mixed contributions when we combine non-Abelian with superconformal generalization. There is only one such mixed term that can arise here. This is from $\delta^2 D^{ij} = ... - (i/2) \bar\epsilon_f \Gamma^P \h\Gamma^{ij} \Gamma^M [\delta' \psi,\sigma] V_M V_P$, where $\delta'\psi = 4\eta \sigma$. But that term is zero by $[\sigma,\sigma] =0$. To the Lagrangian we need to add the term $\frac{1}{2g^2} D^{ij}\cdot D^{ij}$.

\section{Discussion}
We have obtained non-Abelian tensor and hypermultiplet theories with off-shell 6d $(1,0)$ superconformal symmetry where we found that we had to impose the constraint $\L_V = 0$ on all fields. The vector field $V_M$ has the interpretation of a gauge coupling constant, though it is a vector field and not just a constant parameter. It has length dimension one and transforms as a spin-one field under conformal transformations. The vector field is constrained by 
\bea
\L_V \epsilon_f &=& 0\cr
D_M \epsilon_f &=& \Gamma_M \eta
\eea
Our 6d theory might find uses in localization computations. This might be an easier approach in some cases, than performing dimensional reduction to 5d and carrying out localization computations in 5d. For a situation where the fiber length $2\pi g$ is not constant, it may be an advantage to be able to work directly in a 6d covariant formulation. We may keep the 6d symmetries of the problem all the way until we implement the constraint $\L_V = 0$, which is where we break the symmetry, and which is where we may encounter the biggest difficulties in the computation. In practice this may amount to selecting harmonic modes of some Laplace operator which have vanishing $\L_V$, which may be a difficult problem in some cases. However, we expect that one also in 6d will have to consider an instanton factor. Instanton-branes in 6d are instantons located in four transverse directions and extended along the time direction and the vector field $V^M$. By taking these nonperturbative effects into account, we expect that $V^M$ will not be an observable vector field and so one may ask whether implementing $\L_V = 0$ and taking into account nonperturbative instanton sectors, could in fact be equivalent with dropping both the constraint $\L_V = 0$ and all the nonperturbative sectors. If that is true, then that could lead to simpler localization computations directly in 6d. One may reformulate the YM term in terms of a selfdual $H_{MNP}$ and also try to find a Lagrangian whose coupling constant is fixed by selfduality. Such a Lagrangian may be found if one uses for instance the $6=3+3$ formalism \cite{Ko:2013dka}. We may still encounter $V^M$ in cubic interaction terms. However, in localization we  only need to be concerned with the quadratic terms. If those quadratic terms can be rephrased such that they are all indepedendent of $g$ and $V^M$, then the localization computation may become tractable even when $g$ is a field that depends on the spacetime position. The difficult step would then be to implement the constraint $\L_V =0$ on all the fields in the path integral. But as we have speculated above, maybe we do not really need to impose $\L_V = 0$ in the computation of the one-loop determinants, and in that way we could hope to get the nonperturbative contribution for free. Typically the one-loop determinant involves the determinant of a Laplace-like operator. We expect full 6d covariance at the full quantum level. The most naive way to achieve 6d covariance at the quantum level would be to drop $\L_V = 0$ in the perturbative determinants, and at the same time drop the nonperturbative contributions. It remains to be seen whether that naive prescription can give the correct result. To be able to put the 6d theory on a wider class of six-manifolds, one may also couple the 6d theory to supergravity background fields.

\subsubsection*{Acknowledgements} 
This work was supported by the grant Geometry and Physics from Knut and Alice Wallenberg foundation.

\appendix
\section{Appendices}
\subsection{Gamma matrix identities}
For the closure on $\psi$ we have used the identities
\bea
\Gamma^{MNP} \Gamma_R \Gamma_{NP} &=& -24 \delta_R^M + 4 \Gamma_R \Gamma^M\cr
\Gamma^{MNP} \Gamma_{RST} \Gamma_{NP} &=& 4 \Gamma^M \Gamma_{RST}\cr
\Gamma^M \Gamma_R &=& 2\delta^M_R - \Gamma_R \Gamma^M\cr
\Gamma^{MNP} \Gamma_{MNP} &=& -120\cr
\Gamma^{MNP} \Gamma_{RS} \Gamma_{MNP} &=& 24 \Gamma_{RS}\cr
\Gamma^M \Gamma_M &=& 6\cr
\Gamma^M \Gamma_{RS} \Gamma_M &=& 2 \Gamma_{RS}\cr
\Gamma^{MNP} \Gamma_R \Gamma_{MNPQ} &=& 60 g_{RQ} - 12 \Gamma_{RQ}\cr
\Gamma^{MNP} \Gamma_{RST} \Gamma_{MNPQ} &=& 12 \Gamma_Q \Gamma_{RST}\cr
\Gamma^M \Gamma_R \Gamma_{MQ} &=& - 2g_{TQ} - 3 \Gamma_R \Gamma_Q\cr
\Gamma^M \Gamma_{RST} \Gamma_{MQ} &=& \Gamma_Q \Gamma_{RST}
\eea
For the closure on $A_M$ we have used
\bea
\{\Gamma_{MN},\Gamma^{RST}\} &=& - 12 \delta_{MN}^{RS} \Gamma^T + 2 \Gamma_{MN}{}^{RST}
\eea
For the closure on $\psi$ in the Lie algebra formulation, we have used
\bea
\Gamma^{MNP} \Gamma_Q \Gamma_{NR} &=& \Gamma^{MP}{}_{QR} + 3 \Gamma^{MP} g_{QR} - 2\Gamma^M{}_Q \delta^P_R + 2\Gamma^P{}_Q \delta^M_R\cr
&& + 3 \Gamma^M{}_R \delta^P_Q - 3 \Gamma^P{}_R \delta^M_Q - 4 \delta^M_Q \delta^P_R + 4 \delta^P_Q \delta^M_R
\eea

\subsection{The conformal Killing spinor equation}
Here we note some implications of the conformal Killing spinor equation,
\bea
D_M \epsilon_f &=& \Gamma_M \eta
\eea
We define
\bea
[D_M,D_N] \epsilon &=& \frac{1}{4} R_{MNPQ} \Gamma^{PQ} \epsilon\cr
R_{MN} &=& g^{PQ} R_{MPNQ}\cr
R &=& g^{MN} R_{MN}
\eea
We have the identities
\bea
5 D^M D_M \epsilon = \Gamma^{MN} D_M D_N \epsilon = - \frac{R}{4} \epsilon
\eea
They follows from
\bea
\Gamma^N \Gamma^M D_M D_N \epsilon = \Gamma^N \Gamma^M \Gamma_N D_M \eta = - 4 \Gamma^M D_M \eta
\eea
and
\bea
D^M D_M \epsilon = \Gamma^M D_M \eta
\eea
We have 
\bea
\Gamma^M D_M \eta &=& -\frac{R}{20} \epsilon
\eea
and then we get
\bea
\Gamma^M [D_{N},D_{M}] \epsilon &=& 4 D_N \eta - \frac{R}{20} \Gamma_N \epsilon
\eea
so 
\bea
4 D_N \eta &=& \frac{1}{4} R_{NMPQ} \Gamma^M \Gamma^{PQ} \epsilon + \frac{R}{20} \Gamma_N \epsilon
\eea
Finally we use $R_{N[MPQ]} = 0$ and get
\bea
4 D_N \eta &=& - \frac{1}{2} R_{NQ} \Gamma^Q \epsilon + \frac{R}{20} \Gamma_N \epsilon
\eea
We may check the consistency of this relation by contracting by $\Gamma^N$.

\subsection{Conformal actions, Abelian case} 
Here we show that the Abelian actions
\bea
\L_T &=& - \frac{1}{24} H^{MNP} H_{MNP} - \frac{1}{2} D^M \sigma D_M \sigma + \frac{i}{2} \bar\psi \Gamma^M D_M \psi - \frac{R}{10} \sigma^2
\eea
and 
\bea
\L_H &=&  - \frac{1}{2} (D_M \phi^i)^2 + \frac{i}{2} \bar\chi \Gamma^M D_M \chi - \frac{R}{10} \phi^i \phi^i
\eea
are separately superconformal. All that we need to assume here, is that the supersymmetry parameter satisfies
\bea
D_M \epsilon &=& \Gamma_M \eta
\eea

\subsubsection{The tensor multiplet} 
We begin with the Abelian tensor multiplet and split the Lagrangian into two parts, 
\bea
\L_0 &=& -\frac{1}{24} H^{MNP} H_{MNP} - \frac{1}{2} D^M \sigma D_M \sigma + \frac{i}{2} \bar\psi \Gamma^M D_M \psi\cr
\L_1 &=& - \frac{\mu}{2} \sigma^2
\eea
Likewise we split the variation into two parts,
\bea
\delta_0 B_{MN} &=& i \bar\epsilon \Gamma_{MN} \psi\cr
\delta_0 \sigma &=& -i \bar\epsilon \psi\cr
\delta_0 \psi &=& \frac{1}{12} \Gamma^{MNP} \epsilon H_{MNP} + \Gamma^M \epsilon D_M \sigma 
\eea
and 
\bea
\delta_1 \psi &=& \lambda \Gamma^M (D_M \epsilon) \sigma
\eea
We note the following identity
\bea
\bar\psi \Gamma^{MN} D_M \epsilon D_N \sigma + 5 \bar\psi D_M \epsilon D^M \sigma &=& 0
\eea
To prove this identity, we use $D_M \epsilon = \Gamma_M \eta$ to rewrite
\bea
\bar\psi \Gamma^M \Gamma^N D_M \epsilon D_N \sigma &=& - \frac{2}{3} \bar\psi \Gamma^N \Gamma^M D_M \epsilon D_N \sigma
\eea
from which the identity immediately follows. By using this identity, we find 
\bea
\delta_0 \L_0 &=& 4i \bar\psi D_M \epsilon D^M \sigma
\eea
The sign is corresponding to assuming that $\delta$ is anticommuting. Also note that there is no contribution coming from the variation of $H_{MNP}$ thanks to the identity $\Gamma^Q \Gamma_{MNP} \Gamma_Q = 0$. We get
\bea
\delta_0 \L_1 &=& - i \mu \sigma \bar\psi \epsilon
\eea
assuming $\epsilon$ is commuting. Clearly the action is not invariant unless we also consider $\delta_1$ variations. We find 
\bea
\delta_1 \L_0 &=& - 6 i \lambda \bar\psi D_M \epsilon D^M \sigma + i \lambda \frac{3 R}{10} \bar\psi \epsilon \sigma\cr
\delta_1 \L_1 &=& 0
\eea
Thus by taking 
\bea
\lambda &=& \frac{2}{3}\cr
\mu &=& \frac{R}{5}
\eea
we see that the action becomes superconformal. 

\subsubsection{The hypermultiplet}
We decompose the Lagrangian as $\L = \L_0 + \L_1$ where
\bea
\L_0 &=& - \frac{1}{2} (D_M \phi^i)^2 + \frac{i}{2} \bar\chi \Gamma^M D_M \chi\cr
\L_1 &=& - \frac{R}{10} \phi^i \phi^i
\eea
Likewise we decompose the variations as $\delta = \delta_0 + \delta_2$ where
\bea
\delta_0 \phi^i &=& i \bar\epsilon\h\Gamma^i\chi\cr
\delta_0 \chi &=& \Gamma^M \h\Gamma^i \epsilon D_M \phi^i\cr
\delta_1 \chi &=& - 4\h\Gamma^i \eta \phi^i
\eea 
We get the variations
\bea
\delta_0 \L_0 &=& - D^M \phi^i D_M \delta \phi^i - i \bar\chi \Gamma^M D_M \delta \chi\cr
&=& 4 i \bar\chi \h\Gamma^i \Gamma^M\eta D_M\phi^i
\eea
and
\bea
\delta_1 \L_0 &=& - i\bar\chi \Gamma^M D_M \delta_1 \chi\cr
&=& - 4 i \bar\chi \h\Gamma^i\Gamma^M\eta D_M\phi^i - 4 i \bar\chi \h\Gamma^i\Gamma^M D_M\eta \phi^i
\eea
We now need 
\bea
\Gamma^M D_M \eta &=& - \frac{R}{20} \epsilon
\eea
We now get
\bea
\delta \L_0 &=& \frac{R}{5} i \bar\chi \h\Gamma^i \epsilon \phi^i\cr
\delta \L_1 &=& - \frac{R}{5} i \bar\chi \h\Gamma^i \epsilon \phi^i
\eea
and hence $\delta \L = 0$. 

\subsection{Off-shell 10d SYM supersymmetry}
A nice overview of the lower dimensional off-shell supersymmetric theories can be found by starting with the 10d SYM theory \cite{Berkovits:1993hx}, \cite{Pestun:2007rz}. We assume 10d gamma matrices with the chirality matrix $\Gamma^{(10)}$ and spinor and Poincare supersymmetry parameters that are 10d Weyl
\bea
\Gamma^{(10)} \psi &=& \psi\cr
\Gamma^{(10)} \epsilon &=& \epsilon
\eea
and Majorana, $\bar\psi = \psi^T C$ where $C$ can be taken as the same as for our 11d spinor conventions. (In 11d we just include $\Gamma^{(10)}$ as the 11th gamma matrix.) In particular then $C^T = -C$. In this appendix, and only here, we will let $M = 0,1,...,9$. The supersymmetry variations are \cite{Pestun:2007rz}
\bea
\delta A_M &=& \bar\epsilon \Gamma_M \psi\cr
\delta \psi &=& \frac{1}{2} \Gamma^{MN} \epsilon F_{MN} + \nu^{i} K^{i}
\cr
\delta K^{i} &=& -\bar\nu^{i} \Gamma^M D_M \psi
\eea
These variations close off-shell for $\nu^{i}$ being $7$ parameters subject to the constraints
\bea
\bar\epsilon\Gamma_M\nu_{i} &=& 0\cr
\bar\nu^{i} \Gamma_M \nu^{j} &=& - \delta^{ij} \bar\epsilon \Gamma_M \epsilon
\eea
One of the most intricate places in this closure computation, is the term in $\delta^2 K^{i}$ that is bilinear in fermions. This term is given by
\bea
(\delta^2 K^{i})_{bilin} &=& \bar\nu^{i} \Gamma^M [\delta A_M,\psi]\cr
&=& [\bar\psi \Gamma_M \epsilon,\bar\nu^{i}\Gamma^M\psi]
\eea
We now note the Fierz identity 
\bea
\epsilon\bar\nu^{i} &=& c_M \Gamma^M + c_{MNP} \Gamma^{MNP} + c_{MNPQR} \Gamma^{MNPQR}
\eea
The last term drops out because $\Gamma_M \Gamma^{RSTUV}\Gamma^M = 0$. The second term drops out since it is symmetric, $\bar\psi^a \Gamma^{MNP} \psi^b$ is symmetric in the gauge indices $a,b$. Only the first term may contribute, but due to the constraint $\bar\epsilon\Gamma_M \nu^{i} = 0$ the coefficient $c_M$ vanishes itself. Hence $(\delta^2 K^{i})_{bilin} = 0$. 

We can dimensionally reduce on $T^4$ down to 6d where we find the $\N=(1,1)$ vector multiplet. By further imposing the additional Weyl projection on the supersymmetry parameter $\h \Gamma \epsilon = \epsilon$ where $\h\Gamma := \Gamma^{6789}$, we split the 6d (1,1) vector multiplet into one 6d (1,0) vector multiplet and one 6d (1,0) hyper multiplet. Further reduction to 5d gives the following off-shell hypermultiplet supersymmetry variations
\bea
\delta \phi^i &=& \bar\epsilon\h\Gamma^i\chi\cr
\delta \chi &=& \Gamma^{\mu} \h\Gamma^i \epsilon D_{\mu} \phi^i + \Gamma^5 \h\Gamma^i \epsilon [\sigma,\phi^i] + \nu^{\alpha} F^{\alpha}\cr
\delta F^{\alpha} &=& -\bar\nu^{\alpha} \(\Gamma^{\mu} D_{\mu} \chi + \Gamma^5 [\sigma,\chi] + \h\Gamma^i [\phi^i,\psi]\)
\eea
where $\h\Gamma \psi = \psi$ and $\h\Gamma \chi = -\chi$ are vector and hypermultiplet fermions in 5d respectively, and where $\alpha$ now runs over four values for the hyper. Here $\mu = 0,1,2,3,4$ and $\sigma = A_5$ and $\phi^i = A_i$. These variations can be uplifted and related to the 6d variations that we have obtained in the main text, by relating the 10 Weyl projection to the 6d Weyl projection by the following unitary rotation of the spinors,
\bea
\psi &=& U^{\dag} \psi'\cr
\nu^{\alpha} &=& U^{\dag} \nu'^{\alpha}\cr
\epsilon &=& U \epsilon'
\eea
where $U = \frac{1}{\sqrt{2}} (1+ \Gamma^{01234})$. The rotated spinors are no longer subject to the 10d Weyl projections, but instead to the following 6d Weyl projections,
\bea
\Gamma^{01234(10)} \psi' &=& \psi'\cr
\Gamma^{01234(10)} \epsilon' &=& -\epsilon'
\eea
where the M-theory circle (usually taken as the 5th direction) now is along the 10th direction.

\subsection{Many hypermultiplets}
In 5d the hypermultiplet was presented in a different language in \cite{Hosomichi:2012ek} that enables for generalization to $r$ hypermultiplets, labeled by a gauge index $A = 1,2,...,2r$. In this appendix we present the map from our language into their language. But since such a map is available only for $r=1$, we will assume that $r=1$. 

We realize 11d gamma matrices as follows. We split the 11d vector index into $M = 0,1,2,3,4,5$ extended along the M5 brane, and the remaining indices are labeled $A=1,2,3,4,5$. Then we let
\bea
\Gamma^M &=& \Sigma^M \otimes 1\cr
\h\Gamma^A &=& \Sigma \otimes \h\gamma^A
\eea
where the hat does not mean anything. We assume the convention
\bea
\h\gamma^{12345} &=& 1
\eea
We let $M = \{0,m\}$ for $m = 1,2,3,4,5$ and
\bea
\Sigma^0 &=& i \sigma^2 \otimes 1\cr
\Sigma^m &=& \sigma^1 \otimes \gamma^m
\eea
We assume the convention
\bea
\gamma^{12345} &=& 1
\eea
We define the 6d chirality matrices
\bea
\Gamma &=& \Gamma^{012345}\cr
\Sigma &=& \Sigma^{012345}\cr
&=& \sigma^3 \otimes 1\cr
\Gamma &=& \Sigma \otimes 1\cr
\h\Gamma &=& \h\Gamma^{1234}\cr
&=& I \otimes \h\gamma^{1234}
\eea
such that 
\bea
\Gamma \h\Gamma \h\Gamma^5 &=& 1
\eea
We let $A = \{i,5\}$ and we realize
\bea
\h\gamma^i &=& \(\begin{array}{cc}
0 & \sigma^i_{IA}\\
\t\sigma^{iBJ} & 0
\end{array}\)
\eea
where we take
\bea
\sigma^i &=& (\sigma^{1,2,3},-i)\cr
\t\sigma^i &=& (\sigma^{1,2,3},i)
\eea
We then get
\bea
\h\gamma = \(\begin{array}{cc}
-1 & 0\\
0 & 1
\end{array}\) = \(\begin{array}{cc}
-\delta_I^J & 0\\
0 & \delta^A_B
\end{array}\)
\eea
that is acting on a Dirac spinor with the indices 
\bea
\Psi &=& \(\begin{array}{c}
\psi_I\\
\chi^A
\end{array}\)
\eea
We note that for the Pauli sigma matrices we have 
\bea
(\sigma^{1,2,3})^T &=& - \epsilon \sigma^{1,2,3} \epsilon^{-1}
\eea
where
\bea
\epsilon &=& \(\begin{array}{cc}
0 & 1\\
-1 & 0
\end{array}\)
\eea

The 11d charge conjugation matrix is represented as
\bea
C_{11d} &=& \epsilon \otimes C \otimes \h C
\eea
such that 
\bea
C^T &=& -C\cr
\h C^T &=& -\h C\cr
C_{11d}^T &=& - C_{11d}
\eea
We have
\bea
\t \sigma^{iAI} &=& - \epsilon^{IJ} \sigma^i_{JB} \epsilon^{BA}
\eea
We use the conventions
\bea
\h C &=& \(\begin{array}{cc}
\epsilon^{IJ} & 0\\
0 & \epsilon_{AB}
\end{array}\)
\eea
and
\bea
\h C^{-1} &=&  \(\begin{array}{cc}
\epsilon_{IJ} & 0\\
0 & \epsilon^{AB}
\end{array}\)
\eea
where 
\bea
\epsilon^{IJ} = \epsilon^{+-} &=& 1\cr
\epsilon_{AB} = \epsilon_{+-} &=& 1
\eea
and the inverses of these are denoted as
\bea
\epsilon_{IJ} = \epsilon_{+-} &=& -1\cr
\epsilon^{AB} = \epsilon^{+-} &=& -1
\eea
Let us introduce the full index notation of an 11d spinor,
\bea
\Psi &=& \(\begin{array}{c}
\psi^{a\alpha}_I\\
\psi^{a\alpha A}
\end{array}\)
\eea
The M5 brane spinor is 6d Weyl, and decomposes into two tensor multiplet spinors
\bea
\psi^{+\alpha}_I
\eea
labeled by the $SU(2)_R$ index $I = +,-$, and two hypermultiplet spinors
\bea
\chi^{+\alpha A}
\eea
labeled by the $SU(2)_F$ index $A = +,-$. The two supersymmetry parameters are 6d anti-Weyl,
\bea
\epsilon^{-\alpha}_I
\eea
and labeled by the R-symmetry index $I = +,-$. 

We define the Dirac conjugate as
\bea
\bar\epsilon_{+\alpha}^I &=& \epsilon^{-\beta}_J \epsilon_{-+} C_{\beta\alpha} \epsilon^{JI}\cr
\bar\nu_{-\alpha}^I &=& \nu^{+\beta}_J \epsilon_{+-} C_{\beta\alpha} \epsilon^{JI}
\eea
This notation is consistent with the Weyl projections
\bea
\Gamma \epsilon &=& -\epsilon\cr
\bar\epsilon \Gamma &=& \bar\epsilon
\eea
if we represent the 6d chirality matrix when acting on the first index as
\bea
\Gamma = (\sigma^3)^a{}_b
\eea
Namely then,
\bea
(\sigma^3)^-{}_- \epsilon^- &=& - \epsilon^-\cr
\bar\epsilon_+ (\sigma^3)^+{}_+ &=& \bar\epsilon^+
\eea
We define
\bea
\epsilon_{ab} = \epsilon_{+-} = 1
\eea
which means that the Dirac conjugate becomes 
\bea
\bar\epsilon_{+\alpha}^I &=& - \epsilon^{-\beta}_J C_{\beta\alpha} \epsilon^{JI}\cr
\bar\nu_{-\alpha}^I &=& \nu^{+\beta}_J C_{\beta\alpha} \epsilon^{JI}
\eea

We use the gamma matrix realization
\bea
\Gamma^m &=& (\sigma^1)^a{}_b (\gamma^m)^\alpha{}_\beta \(\begin{array}{cc}
\delta_I{}^J & 0\\
0 & \delta^A{}_B
\end{array}\)
\eea
\bea
\h\Gamma^i &=& (\sigma^3)^a{}_b \delta^\alpha{}_\beta \(\begin{array}{cc}
0 & \sigma^i_{IA}\\
\t\sigma^{iBJ} & 0
\end{array}\) 
\eea
Let us now translate
\bea
\delta \phi^i &=& i \bar\epsilon \h\Gamma^i \chi
\eea
into the new language. First, by writing out all indices, we have
\bea
\delta \phi^i &=& - i \bar\epsilon_{+\alpha}^I (\sigma^3)^+{}_+ \delta^{\alpha}_{\beta} \sigma^i_{IA} \chi^{+\beta A}
\eea
that is simplified to 
\bea
\delta \phi^i &=& - i \bar\epsilon_{+\alpha}^I \sigma^i_{IA} \chi^{+\beta A}
\eea
Then we contract both sides with $\t\sigma^{i BJ}$ and use the identity
\bea
\sigma^i_{IA} \t\sigma^{iBJ} &=& 2 \delta^J_I \delta^B_A
\eea
We then get
\bea
\delta \phi^{AI} &=& - 2 i \bar\epsilon_{+\alpha}^I \chi^{+\beta A}
\eea
where we define
\bea
\phi^{AI} &=& \phi^i \t\sigma^{iAI}
\eea

Next
\bea
\delta \chi &=& \Gamma^m \h\Gamma^i \epsilon \partial_m \phi^i
\eea
gives
\bea
\delta \chi^{+\alpha A} &=& - (\gamma^m)^{\alpha}{}_{\beta} \epsilon^{-\beta}_I \partial_m \phi^{AI}
\eea
Let us move on to the auxiliary field. Let us consider the term
\bea
\delta \chi &=& \h\Gamma^i \nu F^i
\eea
Writing out indices, this becomes
\bea
\delta \chi^{+\alpha A} &=& \t\sigma^{iAI} \nu_I^{+\alpha} F^i
\eea
and defining
\bea
F^{AI} &=& t\sigma^{iAI} F^i
\eea
we have
\bea
\delta \chi^{+\alpha A} &=& \nu_I^{+\alpha} F^{AI}
\eea
It is now clear that we can not identify $\nu_I$ with $\epsilon_I$ since they live in opposite 6d Weyl projection spaces. 

Finally we consider
\bea
\delta F^i &=& i \bar\nu \h\Gamma^i \Gamma^m \partial_m \chi
\eea
This becomes
\bea
\delta F^i &=& i \bar\nu_{-\alpha}^I (\gamma^m)^\alpha{}_\beta \sigma^i_{IA} \partial_m \chi^{+\beta A}
\eea
so if we contract both sides by $\t\sigma^{iBJ}$ we get
\bea
\delta F^{AI} &=& 2 i \bar\nu_{-\alpha}^I (\gamma^m)^\alpha{}_\beta \partial_m \chi^{+\beta A}
\eea
Let us now summarize what we have got in the new language,
\bea
\delta \phi^{AI} &=& - 2 i \bar\epsilon^I \chi^{A}\cr
\delta \chi^{A} &=& - \gamma^m \epsilon_I \partial_m \phi^{AI} + \nu_I F^{AI}\cr
\delta F^{AI} &=& 2 i \bar\nu^I \gamma^m \partial_m \chi^{A}
\eea
By remembering how we defined the Dirac conjugates and by noting that our definition differs from \cite{Hosomichi:2012ek} by signs, we can see that we have got an exact agreement with \cite{Hosomichi:2012ek}.

The generalization to non-Abelian case reads
\bea 
\phi^{AI} &=& - 2 i \bar\epsilon^I \chi^{A}\cr
\delta \chi^{A} &=& - \gamma^m \epsilon_I D_m \phi^{AI} + \nu_I F^{AI} - \gamma^m \epsilon_I [v_m,\sigma,\phi^{AI}] - 4 \eta_I \phi^{AI}\cr
\delta F^{AI} &=& 2 i \bar\nu^I \gamma^m D_m \chi^{A} + 2 i \bar\nu^I \gamma^m [\psi_J,v_m,\phi^{AJ}] + 2 i \bar\nu^I \gamma^m [\chi^A,v_m,\sigma]
\eea

\end{document}